\documentclass[english,12pt]{iopart}
\usepackage{float}
\usepackage{iopams}
\usepackage{graphicx}
\usepackage{esint}
\usepackage{color}

\floatstyle{ruled}
\newfloat{algorithm}{tbp}{loa}
\providecommand{\algorithmname}{Algorithm}
\floatname{algorithm}{\protect\algorithmname}

\newcommand{\referee}[1]{{#1}}

\makeatletter
\@ifundefined{showcaptionsetup}{}{%
 \PassOptionsToPackage{caption=false}{subfig}}
\makeatother
\usepackage{subfig}

\usepackage{babel}
\begin{document}

\title[Rare event computation in chaotic systems]{Rare event computation in deterministic chaotic systems using genealogical
particle analysis}
\author{J. Wouters$^{\dag\ddag\star}$\, and F. Bouchet$^\star$}
\address{$\dag$ School of Mathematics and Statistics,\\
University of Sydney, Sydney, Australia}
\address{$\ddag$ Meteorological Institute,\\
University of Hamburg, Hamburg, Germany}
\address{$\star$ Laboratoire de Physique,\\
\'Ecole Normale Sup\'erieure de Lyon, Lyon, France}
\eads{\mailto{jeroen.wouters@uni-hamburg.de}, \mailto{freddy.bouchet@ens-lyon.fr}}


\begin{abstract}
In this paper we address the use of rare event computation
techniques to estimate small over-threshold probabilities of observables
in deterministic dynamical systems. We demonstrate that
genealogical particle analysis algorithms can be successfully applied
to a toy model of atmospheric dynamics, the Lorenz '96 model. We furthermore use
the Ornstein-Uhlenbeck system to illustrate a number of implementation
issues. We also show how a time-dependent objective function based on the fluctuation path
to a high threshold can greatly improve the performance of the estimator
compared to a fixed-in-time objective function.
\end{abstract}

\section{Introduction}

Rare events may have a large impact on the dynamics of geophysical
turbulent flows and the climate. In bistability situations, a rare
transition can drastically change the structure of the flow, like
for instance the bistability of the Kuroshio current \cite{qiu_kuroshio_2000,schmeits_bimodal_2001}
or a change of polarity of the Earth's magnetic field due to the turbulent
dynamics of the Earth metal core \cite{glatzmaier_three-dimensional_1995}.
Rare events can also be extremely important because of their impact
on society, ecosystems or the economy. There are many such examples
in climate dynamics, for example extreme droughts, heat waves, rainfalls
and storms \cite{field_managing_2012}. The probability and the impact
of these events is likely to change in the future due to a changing
climate \cite{field_managing_2012}. The magnitudes of possible changes
are however still uncertain \cite{barriopedro_hot_2011}.

On the one hand, for climate dynamics, there is a lack of sufficient
reliable empirical data \cite{wetter_underestimated_2013}. How could
one assess faithfully the probability of events with recurrence times
longer than one decade with only one or two century long reliable
data? In the last decade, many methods have been developed to extract
the most information possible from this too short time-series. For
instance extreme value statistics \cite{de_haan_extreme_2006,leadbetter_extremes_1983}
has interestingly allowed to extrapolate from the information available
from empirical observation \cite{kharin_estimating_2005,kysely_probability_2002}. 

Another approach would be to critically study and understand the dynamics
of rare events produced by complex climate models. This second approach
seems to be the only available one for events with a recurrence time
longer than decades or centuries. However, the current scientific
state of the art does not yet allow to obtain many results by following
this route. The first critical issue is a
sampling problem. Indeed, if one wants to study events with century
or millennial recurrence time and assess the reliability of the model
dynamics to produce those events, a direct numerical simulation would
require to have model runs of at least hundreds of thousands of years
long in order to get reliable statistics on both the probability of
and the dynamics leading to these events. As it is not always reasonable
to trade computational length with model complexity, especially for
the turbulent part which is responsible for most of the fluctuations,
it is clear that we are facing an extremely difficult scientific challenge.

Is there a way to produce reliable statistics of specific rare events
of a given model, without having to rely on prohibitively long direct
numerical simulations? The same issue has been faced in many other
scientific fields and has led to the development of some interesting
approaches. Indeed, many of the complex systems studied in different
branches of science feature events that are very rare but nevertheless
very relevant due to their high impact. Take for example buffer overflows
in digital communication networks, the insolvency of an insurer or
bank, collisions in planetary systems, the dynamics of phase transitions
in condensed matter, the long time dynamics of complex molecules in
chemistry or biology, to name but a few. In recent years a number
of promising algorithms have been developed to tackle these problems
\cite{rubino_rare_2009,bucklew_introduction_2004,del_moral_mean_2013,moral_feynman-kac_2004}.
These rare event simulation algorithms can drastically reduce the
error made on the estimation of small probabilities.

Generally speaking, the objective of the algorithms is to make rare
events less rare, either by altering the dynamics (importance
sampling) or by targeted killing and cloning of an ensemble of realizations
(genealogical particle analysis or interacting particle algorithms).
Upon estimation the intervention of the algorithm is then taken into
account to obtain an estimate for quantities of the original system. 

Within the class of genealogical particle analysis algorithms, a number
of different strategies exist. A first crucial difference is the type
of quantity one aims at estimating. One can be interested in the distribution
of first entrance time to a set or to sample transition paths \cite{cerou_adaptive_2007,rolland_computing_2015,rolland_statistical_2015,heymann2008geometric},
the probability of a rare event \cite{del_moral_genealogical_2005,garnier_simulations_2006,hairer_improved_2014}
or expectation values of long time averages such as the scaled cumulant
generation function \cite{tailleur_probing_2007,laffargue_large_2013,giardina_simulating_2011,lecomte_numerical_2007,giardina_direct_2006,kurchan_large_2015}
as it appears in large deviation theory. For these different aims
again different algorithms exist, for example geneological algorithms
with fixed \cite{del_moral_genealogical_2005,garnier_simulations_2006}
versus variable \cite{hairer_improved_2014} particle numbers, minimum
action algorithms \cite{heymann2008geometric}, or milestoning \cite{vanden2009exact}.
These algorithms have already been applied to a wide range of systems,
for example percolation problems \cite{adams_harmonic_2008}, in complex
chemistry \cite{chandler_interfaces_2005}, polymer and biomolecule
dynamics \cite{noe_constructing_2009,metzner_illustration_2006,bolhuis_transition_2002,wolde_enhancement_1997},
magnetism \cite{e_energy_2003,kohn_magnetic_2005}, Burger turbulence
\cite{grafke2013instanton,grafke2014arclength}.\\

The aim of this work is to make a first step in the application of
those approaches to climate dynamics problems. Climate dynamics
has specificities that make past approaches not directly adaptable.
First, the climate is clearly out of equilibrium (without
time reversal symmetry or detailed balance), therefore only non-equilibrium
approaches can be considered. Second, the phenomenology of geophysical
turbulent flows is dominated by large scale synoptic scales and is
rather different from other complex dynamics, for instance molecular
dynamics. And third, most climate models are deterministic models,
or sometimes include a stochasticity that does not affect directly
the synoptic scales. 

The aim of this paper is to consider the latter specificity of many
climate model. We address the following question: can rare event algorithms
based on genealogical particle analysis be used effectively and efficiently
for deterministic dynamics? Most algorithms rely on a Markov assumption,
which is verified for deterministic models. However at the cloning
stage, a new trajectory is branched from another one in order to produce
a new ensemble member. For a strictly deterministic system, the offspring
trajectory will not be different from its parent. To ensure separation
of the two trajectories, one has therefore to add either a very small
noise on the overall dynamics or a small change on the initial condition
to the offspring trajectory and rely on the dynamics chaoticity. A
key issue is then to verify a posteriori that the noise is small enough so as
not to distort the measured statistics and probabilities. A test using
different decreasing noise strengths and checking for stability should
therefore be used.

In order to perform the first study of the effectiveness of these
approaches for chaotic deterministic dynamics, we have chosen to study
a simple chaotic system with many degrees of freedom, and of relevance
for climate dynamics: the Lorenz '96 model \cite{lorenz_predictability:_1996,lorenz_designing_2005}.
We have also chosen the conceptually simplest and most robust genealogical
algorithm that allows to sample invariant measure or transition probabilities:
the genealogical particle analysis algorithm. We give a detailed heuristic
presentation of the algorithm and a benchmark on the Ornstein-Uhlenbeck
process in section \ref{sec:ornstein-uhlenbeck}.

The genealogical particle analysis algorithms explore the statistics
of solutions of the dynamical systems by running an ensemble of realizations,
interrupting the ensemble simulation at given times and killing ensemble
members that do not perform well as measured by a weight or objective
function and cloning the ones with a high weight. This selective procedure
explains the terminology genealogical particle analysis. The individual
realizations are also sometimes referred to as particles. The design
of a good objective function is then arguably the main design issue
one faces when using genealogical particle analysis algorithms. Other
choices that have to be made are the number and timing of interactions
and the number of particles to use. We will address these practical
issues in a detailed study of the genealogical particle analysis algorithms
on the Ornstein-Uhlenbeck process. This process is easy to simulate
numerically and allows for analytic expressions to be derived; it
is therefore well suited for the purpose of illustration and testing.

Another aim of this paper is to propose a systematic approach and
procedure to get reliable results and error estimates. We propose
to build the tail of the cumulant distribution funciton of interest by gluing together pieces
of results obtained for different cloning parameter by a systematic
study of the most reliable one, through an empirical estimate of the
algorithm variance. Moreover, we propose a procedure to test empirically
this class of algorithms against the real dynamics. Indeed, for a
model like the Lorenz '96 model, we have no theoretical results that
can serve as a benchmark.\\

The paper is organized as follows. In Section \ref{sec:Rare-event-computation}
we discuss how the need for rare event simulation techniques arises,
what the objective of such algorithms is (making rare events typical)
and how this goal can be achieved for stochastic processes by implementing
a genealogical particle analysis simulation. In Section \ref{sec:Most-probable-fluctuation}
we present a brief discussion of the theory of large deviations and
what it can say about the way in which rare events are reached by
a process. This theory can be used to implement a more efficient rare
event sampling method. In Section \ref{sec:ornstein-uhlenbeck} we
proceed by implementing the genealogical particle analysis simulation
to the Ornstein-Uhlenbeck system. We discuss in depth the selection
of the parameters in the algorithm. In Section \ref{sec:Interacting-particle-algorithm}
we then present the implementation of the genealogical particle analysis
simulation on a chaotic deterministic dynamical system. Finally, we
present our conclusions in Section \ref{sec:Conclusion}.

\section{Rare event computation for Markov dynamics\label{sec:Rare-event-computation}}

In sections \ref{sub:Motivation} and \ref{sub:Importance-sampling}
we present a classical discussion of the inefficiency of brute force
Monte Carlo simulation for estimating small probabilities. This motivates
the need for rare event computation techniques. We introduce the genealogical
particle analysis algorithm and the related theory in Section \ref{sub:Interacting-particle-algorithm}.

\subsection{Motivation\label{sub:Motivation}}

The goal of rare event computation techniques is to make the numerical
estimation of small probabilities more efficient. The necessity of
using such techniques is demonstrated by the sampling of the tail
of a distribution $P$ using independent samples identically distributed
according to the distribution $P$. Say one wants to estimate a small
probability $\gamma_{A}=P(X\in A)\ll1$ by means of a brute force
Monte Carlo estimate 
\begin{eqnarray}
\hat{\gamma}_{A} & = & \frac{1}{N}\sum_{i=1}^{N}1_{A}(X_{i})\label{eq:naiveMC}
\end{eqnarray}
where $1_{A}$ is the indicator function on the set $A$.
The estimator $\hat{\gamma}_{A}$ is an unbiased estimator of $\gamma_{A}$
since the expectation value of $\hat{\gamma}_{A}$ is clearly $\gamma_{A}$.
When the number of samples $N$ is large enough for $\hat{\gamma}_{A}$
to follow a central limit theorem, the statistical error of the estimator
can be quantified by its variance $Var(\hat{\gamma}_{A})=Var(1_{A}(X))/N$.
Furthermore 
\begin{eqnarray}
Var(1_{A}(X)) & = & E((1_{A}(X)-\gamma_{A})^{2})\nonumber \\
 & = & E(1_{A}(X))-\gamma_{A}^{2}=\gamma_{A}-\gamma_{A}^{2}\nonumber \\
 & \approx & \gamma_{A}\label{eq:naiveMCVar}
\end{eqnarray}
when $\gamma_{A}$ is small. The relative error of the estimator $RE$
being proportional to the standard deviation divided by the estimated
quantity, we have $RE\sim\frac{1}{\sqrt{\gamma_{A}N}}$. The relative
error quickly becomes large as $\gamma_{A}$ goes to zero for fixed
sample size $N$. Fortunately there exist methods for estimating small
probabilities more efficiently.

\subsection{Importance sampling\label{sub:Importance-sampling}}

The main ingredient of rare event computation techniques is a sampling
from a modified distribution together with an adapted estimator to
counteract this change of measure. This method to lower the estimator
variance of a rare event probability is termed importance sampling.
Again the example of the sampling from independent identically distributed
random variables provides valuable insights.

Say we want to estimate 
\begin{eqnarray*}
\gamma_{A} & = & \int dX\rho(X)1_{A}(X)\ll1
\end{eqnarray*}
where $\rho$ is the density for our random variable $X$. Instead
of doing a straightforward sampling of $X$ as in (\ref{eq:naiveMC}),
assume we can sample from a modified measure $\tilde{\rho}$ for which
$\tilde{\rho}(X)>0$ whenever $X\in A$ and $\rho(X)>0$. In such
a case, the probability we want to estimate can be rewritten as 
\begin{eqnarray*}
\gamma_{A} & = & \int dX\tilde{\rho}(X)\frac{\rho(X)}{\tilde{\rho}(X)}1_{A}(X)\\
 & = & \tilde{E}(L(X)1_{A}(X))\\
L(X) & := & \frac{\rho(X)}{\tilde{\rho}(X)}\mbox{ whenever }1_{A}(X)\rho(X)>0
\end{eqnarray*}
and we can therefore estimate $\gamma_{A}$ using the estimator

\begin{eqnarray}
\tilde{\gamma}_{A} & = & \frac{1}{N}\sum_{i=1}^{N}L(\tilde{X}_{i})1_{A}(\tilde{X}_{i})\label{eq:importance}
\end{eqnarray}
on samples $\tilde{X}_{i}$ distributed according to $\tilde{\rho}$.

The variance for such an estimator is 
\begin{eqnarray*}
\tilde{V}ar(L(X)1_{A}(X))=\tilde{E}(L^{2}(X)1_{A}(X))-\gamma_{A}^{2} & = & E(L(X)1_{A}(X))-\gamma_{A}^{2}
\end{eqnarray*}
If we could take $\tilde{\rho}$ as the conditional measure with $\tilde{\rho}(X)=\rho(X)/\gamma_{A}$
for $X\in A$ and zero elsewhere, such that $L(X)=1_{A}(X)\gamma_{A}$,
this would result in a zero variance estimator. This estimator is
however not practically implementable, since for this we would need
to know the value of $\gamma_{A}$, which is the value we seek to
calculate.

This calculation demonstrates some important points however. First
of all, it shows that a change of measure can indeed reduce the variance
of the estimator. Although the ideal change of measure is not feasible
in practice, a change of measure that is in a sense close to it should
also give a substantial variance reduction. This modified measure
should therefore have most of its weight on the set of interest $A$.
On the other hand, this also implies that one needs to have some understanding
of the shape of the set $A$ and the distribution on it to construct
an efficient importance sampling.

\subsubsection{Skewing a normal distribution\label{sub:Skewing-a-normal}}

To illustrate how a change of measure can provide significant variance
reductions, even if the modified measure is not the ideal conditional
measure, we discuss an example for normally distributed random variables.
This example will also be useful to illustrate and validate our rare
event algorithm for dynamical systems. 

\begin{figure}
\begin{centering}
\includegraphics[width=0.5\textwidth]{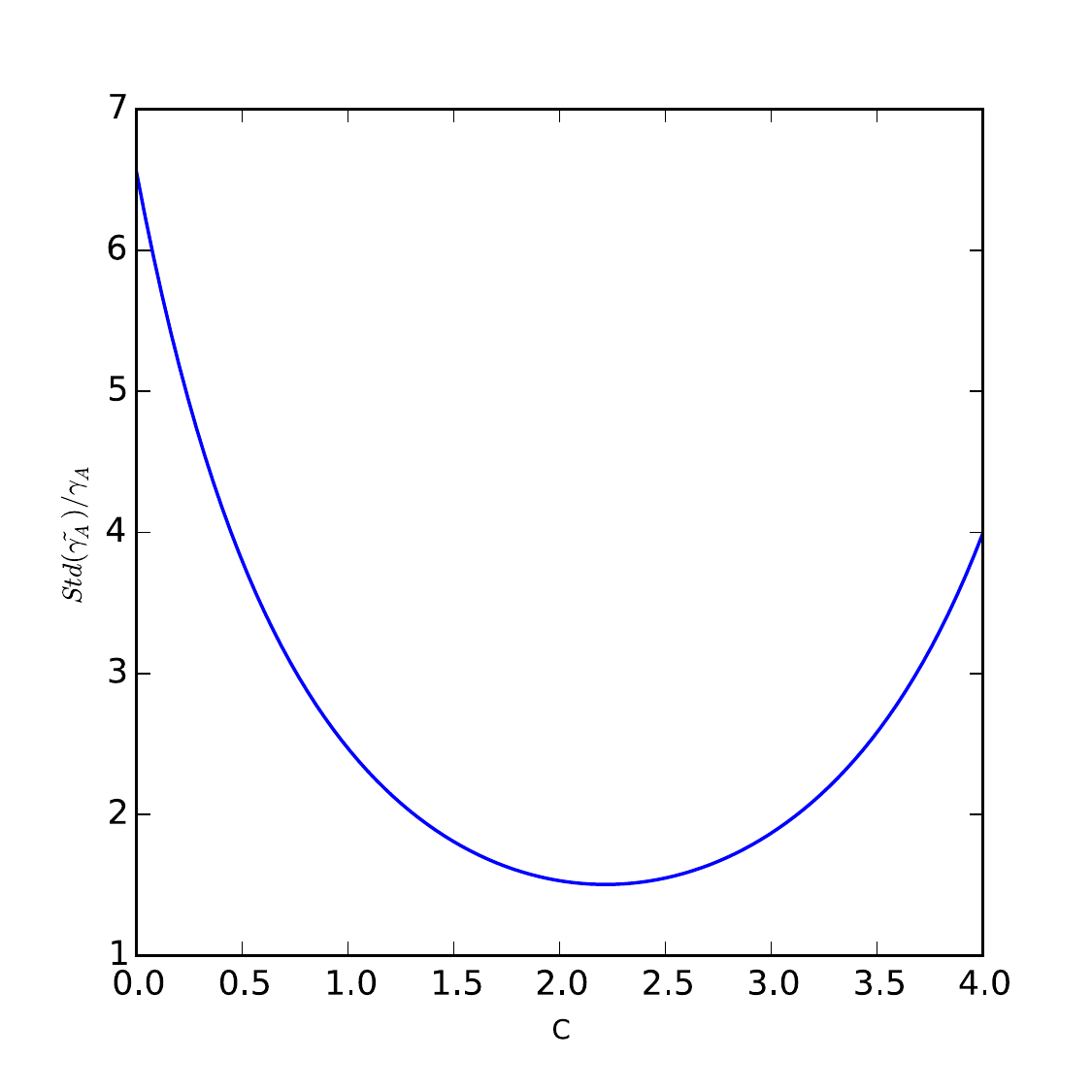}
\par\end{centering}

\caption{The ratio of the standard deviation to estimated probability of an
exponentially tilted gaussian importance sampling estimator for a
threshold $a=2$ with $N=1$\label{fig:gaussian-importance}}
\end{figure}

Say we want to estimate the probability of the rare event $A=\{x>a\}$
for a normally distributed random variable $x\sim\mathcal{N}_{0,1}$
with zero average and standard deviation equal to one. Assume that
we can skew the distribution with an exponential function $\tilde{\rho}(X)=\rho(X)\exp(CX)/E(\exp(CX))=\frac{1}{\sqrt{2\pi}}\exp\left[-\frac{(X-C)^{2}}{2}\right]$
which constitutes of a shift of the average by $C$. Since $L(X)=E\left(e^{CX}\right)/\exp(CX)=\exp\left(-CX+C^{2}/2\right)$,
the variance of the terms in the importance sampling estimator is
now 
\begin{eqnarray}
\tilde{V}ar(L(X)1_{A}(X)) & = & P_{-c,1}(x>a)e^{C^{2}}-\gamma_{A}^{2}\label{eq:gaussian-importance}
\end{eqnarray}
where $P_{\mu,\sigma}$ denotes probabilities under a normal distribution
with mean $\mu$ and variance $\sigma$. The standard deviation, the
square root of the variance (\ref{eq:gaussian-importance}), is plotted
for $a=2$ in Figure (\ref{fig:gaussian-importance}). The standard
deviation has a single minimum, which is obtained for a value of $C$
which is close to $C=2$, for which the mean of the tilted value coincides
with the threshold. This basic example illustrates how importance
sampling can lower dramatically the estimator variance.

As Fig. (\ref{fig:gaussian-importance}) shows, the relative error, which is proportional to the plotted quantity $Std(\tilde{\gamma}_{A})/\gamma_{A}$, can be reduced by a factor of more than $4$ for the case where $A=[2,+\infty)$. Since the error decreases as $1/\sqrt{N}$, this means a at least $16$-fold longer brute force simulation would be necessary to obtain a similarly accurate result. For graphical purposes a relatively low threshold $2$ was chosen here. For higher thresholds, the performance gains increase drastically, with a reduction of computational effort by a factor $6\times10^{5}$ when $A=[5,+\infty)$.

\subsection{Genealogical particle analysis algorithm\label{sub:Interacting-particle-algorithm}}

The motivation for rare event simulation and the discussion of importance
sampling have shown that it is necessary to make rare events less
rare. This concept can be applied to stochastic processes such as
the paths followed by either stochastic dynamics or chaotic deterministic
dynamics. In those cases the objective is to alter the probability
of certain paths that are connected to the rare event one wants to
study.

Two different strategies are employed to alter the path sampling in
stochastic dynamical systems. The first one is to alter the dynamical
equations of the system by introducing a forcing term \cite{vanden-eijnden_rare_2012}. By tuning
a parameter of the added forcing term, one can then attempt to decrease
the variance of the rare event estimator. A second strategy consists
in calculating an ensemble of realizations of the stochastic system
in parallel and manipulating the ensemble members by performing selections at a finite number
of selection times so as to bias the population.

Here we will use the second strategy, by employing a variant of the
so called genealogical particle analysis algorithms. The selections
applied to the ensemble consist of dynamical trajectories, called
particles, being copied and killed depending on weight factor asigned
to every ensemble member. This strategy has the advantage of not altering
the dynamical trajectories themselves, such that their dynamics can
be studied a posteriori.

Extensive analysis of the convergence of genealogical particle analysis
algorithms can be found for example in \cite{del_moral_mean_2013,moral_feynman-kac_2004}.
In the following sections, we perform a simpler calculation, assuming
a mean field approximation, to demonstrate the evolution of the expected
particle distributions in a genealogical particle analysis. The validity
of this mean field approximation for large particle number is the
subject of the complete proofs given in \cite{del_moral_mean_2013,moral_feynman-kac_2004}.
Before going to a truly interacting genealogical particle analysis in Section
\ref{sub:Interacting-particles}, we first get some insight by looking
at an algorithm where particles are reweighted, but by a factor depending
only on the evolution of the particle itself, in Section \ref{sub:A-non-interacting-weighted}.

\subsubsection{A non-interacting genealogical particle analysis \label{sub:A-non-interacting-weighted}}

We calculate rare events of a continuous time Markov chain. $P^{(2)}(y|x,\Delta t)$
denotes the transition probabilities from configuration $x$ to $y$
over a time interval $\Delta t$. We are interested in the probability
of being in a set of configurations $A$ at a time $t=t_{n}$, given
that the process started in configuration $x_{0}$ at time $t=0$. $N_{t}$
denotes the number of particles at time $t$, whereas $\{\xi_{i,t}\}_{1\leqslant i\leqslant N_{t}}$
denote the particle configuration at time $t$. $E_{0,t}$ denotes
expectation values under the original Markov dynamics $P^{(2)}$ at
time $t$. The algorithm to generate the particles is described in
the box “Algorithm \ref{alg:non-interacting-ips}”.

\referee{
The algorithm can be summarized as follows: after initialization (step \ref{alg1:initial}) the ensemble members are evolved forward in time (step \ref{alg1:evolve}) and weight values are calculated from the previous and current configurations of the particle, $\zeta_{i,t_{k}}$ and $\xi_{i,t_{k-1}}$ respectively (step \ref{alg1:non-interacting-weights}). Based on these weight values, ensemble members are killed or cloned (step \ref{alg1:non-interacting-cloning}). Repeating this procedure results in a reweighted sample of paths, from which expectation values of the unweighted path distribution can be estimated (step \ref{enu:non-interacting-final}).
}
The
rare event probability for a set $A$ can be obtained by taking as observable $F(x)=1_{A}e^{-V(x)}e^{V(x_{0})}$
such that $E_{0,t_{n}}(Fe^{V})e^{-V(x_{0})}=E_{0,t_{n}}(1_{A})$.

Note that the random number $N_{i,k}$ generated in step \ref{alg1:non-interacting-cloning}
can be zero, such that particles can be killed as well as cloned (when
$N_{i,k}>1$). A way to generate the random number described in step
\ref{alg1:non-interacting-cloning} is to take $N_{i,k}=\left\lfloor W_{i,k}+u\right\rfloor $
where $u$ is uniformly distributed on $\left[0,1\right]$ and $\left\lfloor x\right\rfloor $
is the floor of $x$ (the largest integer smaller than $x$).

\textbf{\emph{}}
\begin{algorithm}
\begin{enumerate}
\item Initiate $M$ particles in configuration $x_{0}$: $\xi_{i,0}=x_{0}\mbox{ for }1\leqslant i\leqslant N_{t_0}=M$ \label{alg1:initial}
\item For every time step $k\in\{1,\ldots,n\}$

\begin{enumerate}
\item Propagate $\xi_{i,t_{k-1}}$ under the dynamics, resulting in $\zeta_{i,t_{k}}$
distributed according to $P^{(2)}(\zeta_{i,t_{k}}|\xi_{i,t_{k-1}},\Delta t_{k})$
with $\Delta t_{k}=t_{k}-t_{k-1}$ \label{alg1:evolve}
\item Calculate weights $W_{i,k}$ for particle $i$:\label{alg1:non-interacting-weights}
\begin{eqnarray*}
W_{i,k} = W(\zeta_{i,t_{k}},\xi_{i,t_{k-1}}) & \mathrel{\mathop:}= & \exp(V(\zeta_{i,t_{k}})-V(\xi_{i,t_{k-1}}))
\end{eqnarray*}
for a suitably chosen weight function $V$
\item \label{alg1:non-interacting-cloning}Generate a new particle distribution
$\xi_{j,t_{k}}$ consisting of $N_{i,k}$ copies of particle with
configuration $\zeta_{i,t_{k}}$where $N_{i,k}$ is chosen at random
such that $E(N_{i,k})=W_{i,k}$ (note that $N_{t_{k}}=\sum_{i}N_{i,k}$)
\end{enumerate}
\item Finally, for any $F$, calculate $\breve{F}=\frac{1}{M}\Sigma_{i=1}^{N_{t_{n}}}F(\xi_{i,t_{n}})$
to estimate $E_{0,t_{n}}(Fe^{V})e^{-V(x_{0})}$ (to estimate $\gamma_{A}$
take $F(x)=F_A(x):=1_{A}(x)\exp(V(x_{0})-V(x))$)\label{enu:non-interacting-final}
\end{enumerate}
\textbf{\emph{\caption{\textbf{\emph{\label{alg:non-interacting-ips}Non-interacting weighted
particle system}}}
}}
\end{algorithm}

\subsubsection{Unbiased estimator}

We first show that Algorithm \ref{alg:non-interacting-ips} provides
an unbiased estimator for the quantity $E_{0,t_n}(Fe^{V})e^{-V(x_{0})}$,
i.e. the algorithm results in a random estimate whose expectation value equals
the quantity to be estimated:\textbf{\emph{ 
\begin{eqnarray*}
E_{1}\left(\frac{1}{M}\Sigma_{i=1}^{N_{t_{n}}}F(\xi_{i,t_{n}})\right) & = & E_{0,t_{n}}(Fe^{V})e^{-V(x_{0})}
\end{eqnarray*}
}}where $E_{1}$ is the expectation over the random variables in the
algorithm.

Write $N(x,t)$ the particle number at configuration $x$, i.e. $N(x,t)dx$
is the number of particles with $x \leqslant x_{i,t} < x+dx$:
\begin{eqnarray*}
N(x,t_{k-1}) & = & \Sigma_{i=1}^{N_{t_{k-1}}}\delta(x-\xi_{i,t_{k-1}})
\end{eqnarray*}
According the algorithm \ref{alg:non-interacting-ips}, if a particle
sits at $\xi_{i,t_{k-1}}$ at time step $k-1$, $N_{i,k}$ copies
are created of $\zeta_{i,t_{k}}$ at the next time step. Hence, the
particle number at the next time step will be 
\begin{eqnarray}
N(x,t_{k}) & = & \Sigma_{i=1}^{N_{t_{k-1}}}N_{i,k}\delta(x-\zeta_{i,t_{k}})\label{eq:particle-distr-update}
\end{eqnarray}
One step in the algorithm involves the generation of two sets of random variables,
the updated particle configurations $\zeta_{i,t_{k}}$, which is conditioned
on $\xi_{i,t_{k-1}}$, and the number of particle copies $N_{i,t_{k}}$,
which depends on both $\zeta_{i,t_{k}}$ and $\xi_{i,t_{k-1}}$. The
expectation value of functions depending on the particle configurations
$\xi_{i,t_{k}}$ at step $k$ can therefore be expressed as the expectation
value 
\begin{eqnarray*}
E_{\xi_{i,t_{k}}}(\bullet) & = & E_{\xi_{i,t_{k-1}}}(E_{\zeta_{i,t_{k}}|\xi_{i,t_{k-1}}}(E_{N_{i,k}|\zeta_{i,t_{k}},\xi_{i,t_{k-1}}}(\bullet)))
\end{eqnarray*}
Applying this expression to Eq. \ref{eq:particle-distr-update} and
using the probabilities for the updated particle configurations $P(\zeta_{i,t_{k}}|\xi_{i,t_{k-1}})=P^{(2)}(\zeta_{i,t_{k}}|\xi_{i,t_{k-1}},\Delta t_{k})$
and that the number of particle copies $E_{N_{i,{k}}|\zeta_{i,t_{k}},\xi_{i,t_{k-1}}}(N_{i,k})=W_{i,k}(\zeta_{i,t_{k}},\xi_{i,t_{k-1}})$,
we have 
\begin{eqnarray*}
E_{1}(N(x,t_{k})) & = & E_{\xi_{i,t_{k-1}}}(E_{\zeta_{i,t_{k}}|\xi_{i,t_{k}}}(\Sigma_{i=1}^{N_{t_{k-1}}}W_{i,k}(\zeta_{i,t_{k}},\xi_{i,t_{k-1}})\delta(x-\zeta_{i,t_{k}})))\\
 & = & E_{\xi_{i,t_{k-1}}}\left(\Sigma_{i=1}^{N_{t_{k-1}}}\int dy P^{(2)}(y|\xi_{i,t_{k-1}},\Delta t_k)W_{i,k}(y,\xi_{i,t_{k-1}})\delta(x-y)\right)\\
 & = & E_{\xi_{i,t_{k-1}}}\left(\Sigma_{i=1}^{N_{t_{k-1}}}P^{(2)}(x|\xi_{i,t_{k-1}},\Delta t_k)W_{i,k}(x,\xi_{i,t_{k-1}})\right)\\
 & = & E_{\xi_{i,t_{k-1}}}\left(\int dzP^{(2)}(x|z,\Delta t_k)W(x,z)\Sigma_{i=1}^{N_{t_{k-1}}}\delta(z-\xi_{i,t_{k-1}})\right)\\
 & = & \int dzP^{(2)}(x|z,\Delta t_k)W(x,z)E_{1}(N(z,t_{k-1}))\\
 & = & \int dzP^{(2)}(x|z,\Delta t_k)e^{V(x)-V(z)}E_{1}(N(z,t_{k-1}))
\end{eqnarray*}
This equation relates the expected particle density at step $k$ to
the density at step $k-1$. By iteration we can relate the density
at step $k$ to the density at the start of the algorithm, which is
$M\delta(x-x_{0})$: 
\begin{eqnarray*}
E_{1}(N(x,t_k)) & = & \int dx_{k-1}\ldots dx_{1} dz P^{(2)}(x|x_{k-1},\Delta t_k)\ldots P^{(2)}(x_{1}|z,\Delta t_k) \\ && \phantom{\int dx_{k-1}\ldots dx_{1} dz P^{(2)}} \times e^{V(x)-V(z)}M\delta(z-x_{0})\\
 & = & M P^{(2)}(x|x_{0},t_k - t_0)e^{V(x)-V(x_{0})}
\end{eqnarray*}
The expectation value of the quantity calculated at the end of the
algorithm in step \ref{enu:non-interacting-final} is therefore 
\begin{eqnarray*}
E_{1}\left(\frac{1}{M}\Sigma_{i=1}^{N_{t_n}}F(\xi_{i,t_n})\right) & = & \frac{1}{M} E_{1}\left(\int dx\Sigma_{i=1}^{N_{t_n}} \delta(x-\xi_{i,t_n}) F(x)\right)\\
 & = & \frac{1}{M}\int dxE_{1}(N(x,t_n))F(x)\\
 & = & \int dx F(x)e^{V(x)-V(x_{0})}P^{(2)}(x|x_{0},t_n - t_0)\\
 & = & E_{0,t_n}(F e^{V})e^{-V(x_{0})}
\end{eqnarray*}

Note that the expected total particle number $E_{1}(N(t_k))=\int dxE_{1}(N(x,t_{k}))=M\int dxP^{(2)}(x|x_{0},t_k - t_0)e^{V(x)-V(x_{0})}$
is in general not preserved over time. The particle number can strongly
increase, which entails a large numerical cost. The solution to this
problem is to renormalize the weights calculated in step \ref{alg1:non-interacting-weights}
of the algorithm, hence introducing an interaction between the ensemble
members. We discuss this new algorithm in the next section. As we
will see, the interaction complicates the algorithm analysis.

\subsubsection{Interacting particles \label{sub:Interacting-particles}}

We now add interaction to the weights of the particle system, so as
to control the particle number. A similar analysis as in the previous
section can still be carried out, if one assumes that the number of
particles used in algorithm is large enough, such that averages over
particle configurations can be replaced by an expectation value under
the law of large numbers (mean field approximation). The corresponding
algorithm is described in the box “Algorithm \ref{alg:Interacting-particle-system}”.
By applying the algorithm to a function $F_A(x)=1_{A}(x)\exp(V(x_{0})-V(x))$
with $1_{A}$ the indicator function of the set $A$, estimates $\breve{\gamma}_{A}$
of the probability $\gamma_{A}$ can be obtained.

\textbf{\emph{}}
\begin{algorithm}
\begin{enumerate}
\item Initiate $M$ particles in configuration $x_{0}$, $\xi_{0}^{i}=x_{0}\mbox{ for }1\leqslant i\leqslant N_{0}=M$
\item For every time step $k\in\{1,\ldots,n\}$

\begin{enumerate}
\item Propagate $\xi_{i,t_{k-1}}$ under the dynamics, resulting in $\zeta_{i,t_{k}}$
distributed according to $P^{(2)}(\zeta_{i,t_{k}}|\xi_{i,t_{k-1}},\Delta t_{k})$
with $\Delta t_{k}=t_{k}-t_{k-1}$
\item Calculate weights for particle $i$: 
\begin{eqnarray*}
\bar{W}_{i,k} & = & \frac{W_{i,k}(\zeta_{i,t_{k}},\xi_{i,t_{k-1}})}{Z_{k}}\\
Z_{k} & = & \frac{1}{N_{t_{k}}}\Sigma_{i}W_{i,k}(\zeta_{i,t_{k}},\xi_{i,t_{k-1}})
\end{eqnarray*}

\item Store the value of the normalizing factor $Z_{k}$
\item \label{enu:interacting-cloning}Generate a new particle distribution
$\xi_{j,t_{k}}$ consisting of $N_{i,k}$ copies of particle with
configuration $\zeta_{i,t_{k}}$where $N_{i,k}$ is chosen at random
such that $E(N_{i,k})=\bar{W}_{i,k}$
\end{enumerate}
\item Finally calculate $\frac{1}{M}\Sigma_{i=1}^{N_{t_{n}}}F(\xi_{i,t_{n}})\prod_{k=1}^{n}Z_{k}$
to estimate $E_{0}(Fe^{V})e^{-V(x_{0})}$ (for $\gamma_{A}$ take
$F(x)=1_{A}(x)\exp(V(x_{0})-V(x))$)
\end{enumerate}
\textbf{\emph{\caption{\textbf{\emph{\label{alg:Interacting-particle-system}Interacting
genealogical particle analysis}}}
}}
\end{algorithm}

\begin{figure}
\begin{centering}
\includegraphics[scale=0.75]{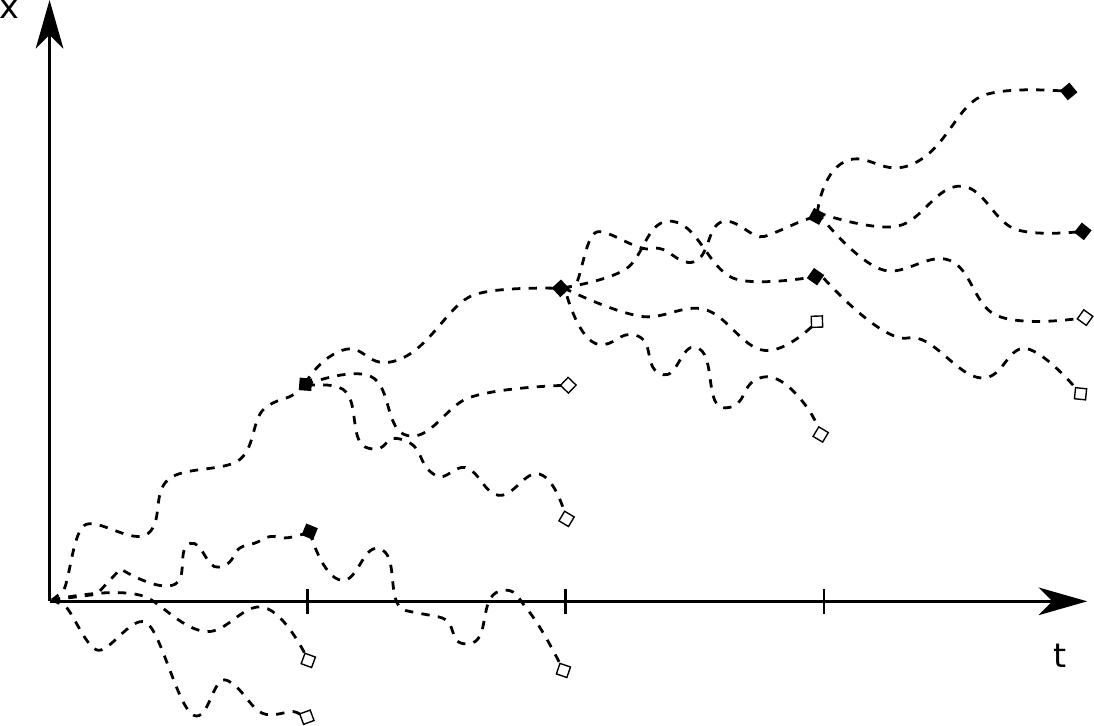}
\par\end{centering}

\caption{Schematic representation of the genealogical particle analysis}
\end{figure}
We again perform an analysis of the evolution of the expected
particle distribution in Algorithm \ref{alg:Interacting-particle-system}.
For simplicity of the derivation we assume that the number of particles
$N_{t_k}$ in the algorithm is large, such that by the law of
large numbers, 
\begin{eqnarray}
Z_{k}=\frac{1}{N_{t_{k}}}\Sigma_{i}^{N_{t_{k}}} W(\zeta_{i,t_{k}},\xi_{i,t_{k-1}}) & \approx & E_1(W(\zeta_{i,t_k}, \xi_{i,t_{k-1}})) \label{eq:normalization-factor}\\
 & = & \int dx \, dy\frac{E_1(N(x,t_{k-1}))}{E_1(N_{t_{k-1}})}P^{(2)}(y|x,\Delta t_k)W(y,x)
\end{eqnarray}
Using this estimate and the same reasoning as for the non-interacting
particle algorithm, we have for the expected particle distribution
that 
\begin{eqnarray}
E_1(N(x,t_{k})) & \approx & \int dz P^{(2)}(x|z,\Delta t_k)\frac{W(x,z)}{E_1(W)}E_1(N(z,t_{k-1}))\label{partdens3}
\end{eqnarray}
The expectation value of $W$ in the denominator can be substituted
using Eq. \ref{eq:normalization-factor}. The particle number is now
constant: 
\begin{eqnarray*}
E_1(N_{t_{k}})) & = & \int dx E_1(N(x,t_{k}))=E_1(N_{t_{k-1}})=\ldots=E_1(N_{t_0})=M
\end{eqnarray*}
We therefore have that 
\begin{eqnarray*}
E_1(W(\zeta_{i,t_k}, \xi_{i,t_{k-1}})) & \approx & \int dxdyP^{(2)}(x|y, \Delta t_k)W(x,y)\frac{E(N(x,t_{k-1}))}{M}
\end{eqnarray*}
Inserting this into \ref{partdens3}, we have 
\begin{eqnarray*}
E(N(x,t_{k})) & \approx & M\frac{\int dyP^{(2)}(x|y, \Delta t_k)W(x,y)E(N(y,t_{k-1}))}{\int dxdyP^{(2)}(x|y, \Delta t_k)W(x,y)E(N(y,t_{k-1}))}
\end{eqnarray*}
Therefore by iteration 
\begin{eqnarray*}
\frac{1}{M}\Sigma_{i=1}^{N}F(x_{i,t_{n}}) & \approx & 
\frac{E_0(F(x_{n})W(x_{n},x_{n-1})\ldots W(x_{1},x_{0}))}{E_0(W(x_{n},x_{n-1})\ldots W(x_{1},x_{0}))} \\ &=&\frac{E_0(F(x_{n})W(x_{n},x_{n-1})\ldots W(x_{1},x_{0}))}{\mathcal{Z}_{n}}
\end{eqnarray*}
where $\mathcal{Z}_{n}=E_0(W(x_n,x_{n-1}) \ldots W(x_1,x_0))$. From Eqs. \ref{eq:normalization-factor} and \ref{partdens3} we see that $Z_k \approx \frac{\mathcal{Z}_k}{\mathcal{Z}_{k-1}}$ and therefore $\frac{1}{M}\Sigma_{i=1}^{N}F(x_{i,t_{n}}) \prod_{k=1}^n Z_k \approx E_0(F(x_{n})W(x_{n},x_{n-1})\ldots W(x_{1},x_{0})) = E_{0}(Fe^{V})e^{-V(x_{0})}$

The above reasoning can also be extended to show that path dependent quantities
(such as $E[x(\tau)|x(T)>a]$ for $\tau<T$) can be estimated from
the ancestral paths of the particle system.

\subsubsection{Time-dependent weighting\label{sub:Time-dependent-weighting}}

The weighting function $W(x,y)=\exp(V(x)-V(y))$ results in a particle
distribution tilted by $\exp(V(x))$ at all selection times $t_k$. More flexibility
can be obtained by using time-dependent weighting, for example with
a weighting function of the form
\begin{equation}
W(t_k,x,y)=\exp(V_{t_k}(x)-V_{t_{k-1}}(y))\label{eq:time-dep-weight}
\end{equation}

This way the telescoping canceling of $V_{t_k}$  is preserved in products of weights
that appear in the calculation of the tilted measure. For example,
\begin{eqnarray*}
W(t_k,x,y)W(t_{k-1},y,z) & = & e^{V_{t_k}(x)-V_{t_{k-1}}(y)}e^{V_{t_{k-1}}(y)-V_{t_{k-2}}(z)}\\
 & = & \exp(V_{t_k}(x)-V_{t_{k-2}}(z))
\end{eqnarray*}
The result is again a particle distribution tilted by $\exp(V_{t_k}(x))$
at time $t_k$, as with the time-independent weight function. However,
paths up to the final time will have different weights, which can
make a large difference in the algorithm performance, as we will demonstrate
in Section \ref{sub:Selections-along-the}.

\section{Fluctuation paths and the weighting function\label{sec:Most-probable-fluctuation}}

The ideal change of measure discussed in Section \ref{sub:Importance-sampling}
suggests to make the rare event that is the least rare the most probable
one under the reweighted dynamics. This rationale extends not only
to the distribution of the system at the final time, but also to the
entire path up to the final time. This means that variance can be
reduced if the least unlikely path leading to a high threshold is
made more likely under the particle system dynamics. 

For stochastic differential equations in the weak noise limit, the
least unlikely path from an attractor can be calculated from Freidlin-Wentzell
type large deviation theory and is called a fluctuation path (also
sometimes an instanton). The particle system dynamics can be made
to more closely follow the fluctuation path by using the time-dependent
weighting discussed in Section \ref{sub:Time-dependent-weighting}. Even if the particle distribution at the final time
is the same as the particle distribution obtained with a constant weighting, there
is still a variance reduction since less particles are killed, increasing
the independence of the particle and thus the effective particle number.

\subsection{Fluctuation paths\label{sub:Fluctuation-paths}}

The probability of a given path in a stochastic differential equation
with small noise,
\[
dX^{\epsilon}=b(X^{\epsilon})dt+\sqrt{\epsilon}dW,
\]
where $W$ is a Brownian motion, can be estimated using the Freidlin-Wentzell large deviation theory.
The theory determines the probability of seeing a path that is close
to a specified continuous function in the limit of $\epsilon$ going
to zero. It roughly states that
\[
\lim_{\epsilon\rightarrow0}\epsilon\log P[X^{\epsilon}\in F]=-\inf_{\omega\in F}I(\omega)
\]
where $F$ is any closed subset of the set of continuous trajectories
and the rate functional $I$ is called the action. The action is given
by
\begin{eqnarray}
I(\omega)=\frac{1}{2}\int_{0}^{T}dt(\dot{\omega}(t)-b(\omega(t)))^{2}=\int_{0}^{T}dt\mathcal{L}[\omega,\frac{\partial\omega}{\partial t}]\label{eq:Action} \\
\mathcal{L}[\omega,\frac{\partial\omega}{\partial t}] := \frac{1}{2} (\dot{\omega}(t)-b(\omega(t)))^{2} 
\end{eqnarray}

The distribution of paths leading to rare fluctuations then concentrates
around action minima as $\epsilon$ decreases, with given constraints.
If the set of paths $F$ contains the evolution along the deterministic
dynamics $\dot{x}=b(x)$, this path will obviously minimize the above
action, hence the need for constraints to obtain more interesting
results. For example, in the simple case were the deterministic dynamics
$\dot{x}=b(x)$ has a single attractor $x_{0}$, the distribution
of the paths conditioned on $X(0)=X_0, \, X_0 \neq x_0$ concentrate close to the minima
of the action $\int_{-\infty}^{0}dt\mathcal{L}[\omega,\frac{\partial\omega}{\partial t}]$
with the boundary conditions $X(-\infty)=x_{0}$ and $X(0)=X_0$. Such
a path is called a fluctuation path leading to $X_0$ (it is also sometimes
called an instanton, but instanton usually rather refers to those
fluctuation paths that connect attractors to saddle points).

\section{Rare event simulation for a stochastic process: the Ornstein-Uhlenbeck
process}

\label{sec:ornstein-uhlenbeck}

We now illustrate some of the practical issues arising when implementing
a genealogical particle analysis algorithm for rare event estimation.
We start off with a stochastic process for which we can calculate
explicitly all of the probabilities that we want to estimate, for
pedagogical reasons, and so that we can compare the numerical results
to the analytic expressions.

\subsection{Description of the Ornstein-Uhlenbeck process}

We consider the Ornstein-Uhlenbeck process
\begin{equation}
dx=-\lambda xdt+\sigma dW\label{eq:ornstein-uhlenbeck}
\end{equation}
As the transition probabilities $P(x(t)|x(0))$ are Gaussian, the Ornstein-Uhlenbeck process preserves Gaussianity. Using
the It\^o formula, one can derive that the mean $m(t)=E(x(t))$ and
the variance $v(t)=E((x(t)-m(t))^{2})$ evolve according to the equations
\begin{eqnarray}
\dot{m} & = & -\lambda m\nonumber \\
\dot{v} & = & \sigma^{2}-2\lambda v\label{eq:v}
\end{eqnarray}
which can be easily solved explicitly.

The probability that $x$ exceeds a certain threshold $a$ at a time
$t$, given that the process started at $x(0)=0$ at time zero can
be calculated explicitly as
\begin{equation}
P(x(t)>a|x(0)=0)=\int_{a}^{\infty}dx \mathcal{N}_{m(t),v(t)}(x),\label{eq:Pa}
\end{equation}
where $m$ and $v$ solve (\ref{eq:v}) and $\mathcal{N}$ is the probability density function of the normal distribution with mean $m$ and variance $v$. We consider in the following
the estimation of this probability through a genealogical particle
analysis algorithm.
\referee{Below we will use the parameter values $\lambda=1$ and $\sigma=1$.}

\subsection{Algorithm implementation}

Let us assume we seek to estimate a small probability $\gamma_{A}$,
for instance $\gamma_{A}=P(x(t)\in A|x(0)=0)$. We denote $M$ the
number of particles for each realization of the algorithm. Then each
independent realization $i$ of the algorithm, with $M$ particle
each, will give an estimate $\breve{\gamma}_{A,i}$. According to
a theorem discussed in \cite{del_moral_genealogical_2005}, asymptotically
for large $M$, the random number $\breve{\gamma}_{A,i}$ is distributed
according to a Gaussian distribution with standard deviation $\sigma_{A}(M)=\sigma_{A}/\sqrt{M}$
and a corresponding relative error $RE(M)=\sigma_{A}(M)/\gamma_{A}$.
The value of the estimator relative error $RE(M)$ is essential as
it quantifies the relative error one should expect for each realization
of the algorithm, and thus the quality of the result. How the estimator
relative error $RE(M)$ depends on the number of selections, on their
timing, and the type observables are critical questions that we analyze
in this section.

\subsubsection{Number of particles}

The result in \cite{del_moral_genealogical_2005} proves the existence
of the central limit theorem, but does not give a value for the estimator
variance $RE(M)$. In order to get an estimate of $RE(M)$, we compute
it empirically by performing $K$ independent algorithm realization
and using the estimator
\begin{equation}
RE(M)\simeq\sqrt{\frac{1}{K}\sum_{i=1}^{K}(\breve{\gamma}_{A,i}-\gamma_{A})^{2}}/\gamma_{A}.\label{eq:Estimated_RE}
\end{equation}
In this formula the value of $\gamma_{A}$ will be either the theoretical
value when it is available, for instance for the Ornstein-Uhlenbeck
process, or the estimated value of the probability by averaging $\breve{\gamma}_{A,i}$
over $K$ realizations. In the following, by an abuse of notation,
$RE$ denotes either the theoretical estimator variance of the estimator
variance evaluated from (\ref{eq:Estimated_RE}), which one should
be clear from the context. \\

\begin{figure}
\subfloat[]{\begin{centering}
\includegraphics[width=0.45\textwidth]{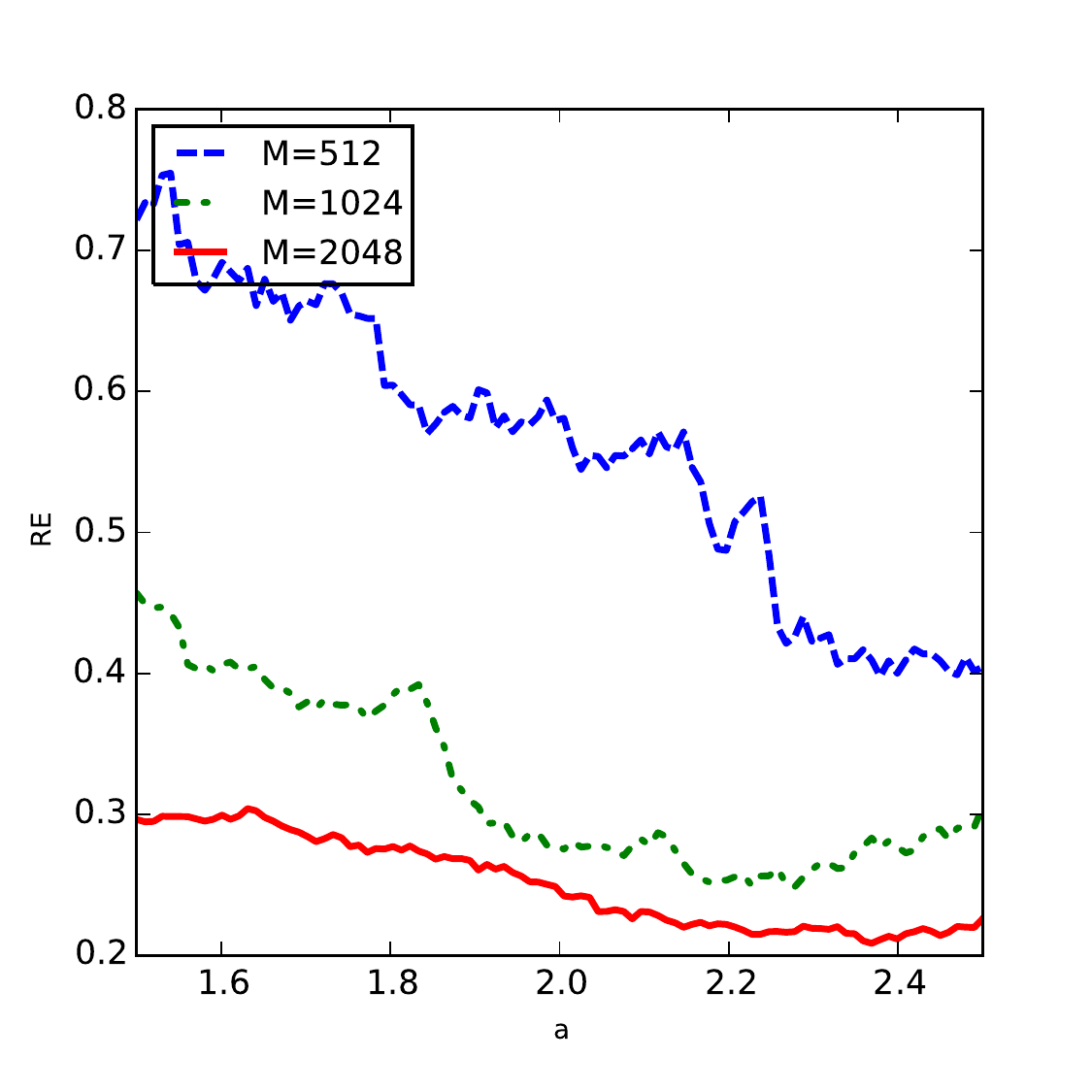}\label{fig:numparta}
\par\end{centering}

}\ \subfloat[]{\begin{centering}
\includegraphics[width=0.45\textwidth]{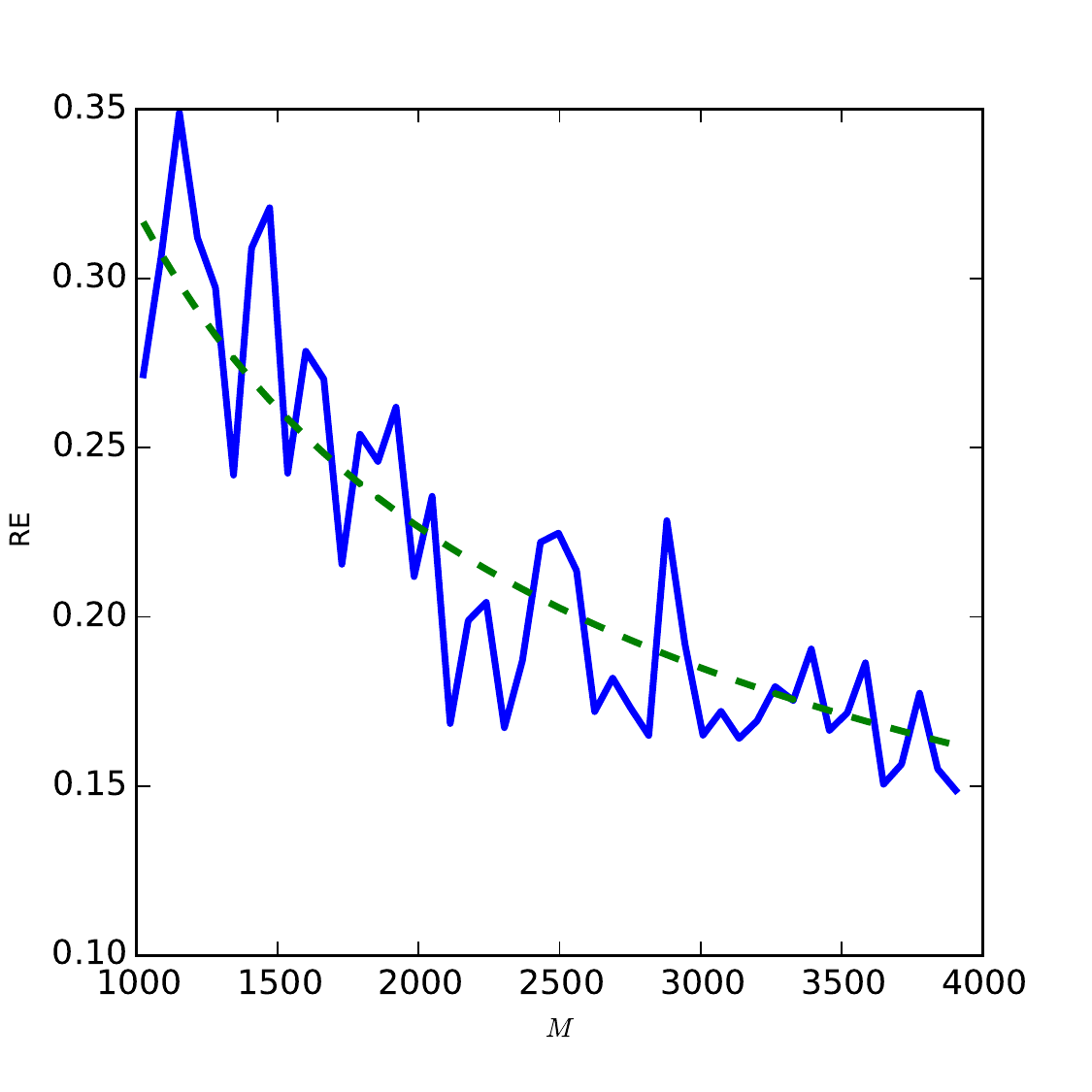}\label{fig:numpartb}
\par\end{centering}

}\caption{(a) The estimator variance $RE=\sqrt{\frac{1}{K}\sum_{i=1}^{K}(\breve{\gamma}_{A,i}-\gamma_{A})^{2}}/\gamma_{A}$
where $A=[a,+\infty)$ for different values of the threshold $a$
and number of particles $M$. $RE$ is estimated from $K=50$ independent
runs of the genealogical particle analysis algorithm. The weight function
is $\exp(C\Delta x)$ with $C=4$. For comparison: the brute force
Monte Carlo estimator variance with $512$ particles for a threshold
$a=2$ is $0.95$. (b) The estimator variance for a fixed threshold
$a=2$ for different numbers of particles $M$. A $1/\sqrt{M}$ function
is fitted and shown as the dashed line.\label{fig:numpart}}
\end{figure}

We first study the estimator variance $RE$ for the Ornstein–Uhlenbeck
case. We first test whether or not the regime of the central limit
theorem has been reached by changing the number of particles $M$,
and verifying whether $RE$ (\ref{eq:Estimated_RE}) reduces by the
corresponding $\sqrt{M}$ factor. Figure \ref{fig:numparta} shows
the expected decrease in relative error as the number of particles
is increased for a range of thresholds $a$ (see eq. (\ref{eq:Pa})).
The inverse square root behavior of the error with increasing number
of particles is demonstrated for a fixed threshold in Figure \ref{fig:numpartb}.
The parameters are specified in the figure captions. 

We study how the estimator variance $RE$ depends on the other numerical
parameters in the following sections.

\subsubsection{Number of selections and their timing}

Since little theoretical analysis has been performed on the optimal
number of selection steps, this is the most heuristic choice to be
made. Some numerical analysis of this issue has been performed in
\cite{el_makrini_diffusion_2007}. For the problem they investigate,
changing the number of selections, and using equidistant in time selections,
the estimator variance clearly shows a minimum for a certain number
of selections.

\begin{figure}
\subfloat[]{\centering{}\includegraphics[width=0.45\textwidth]{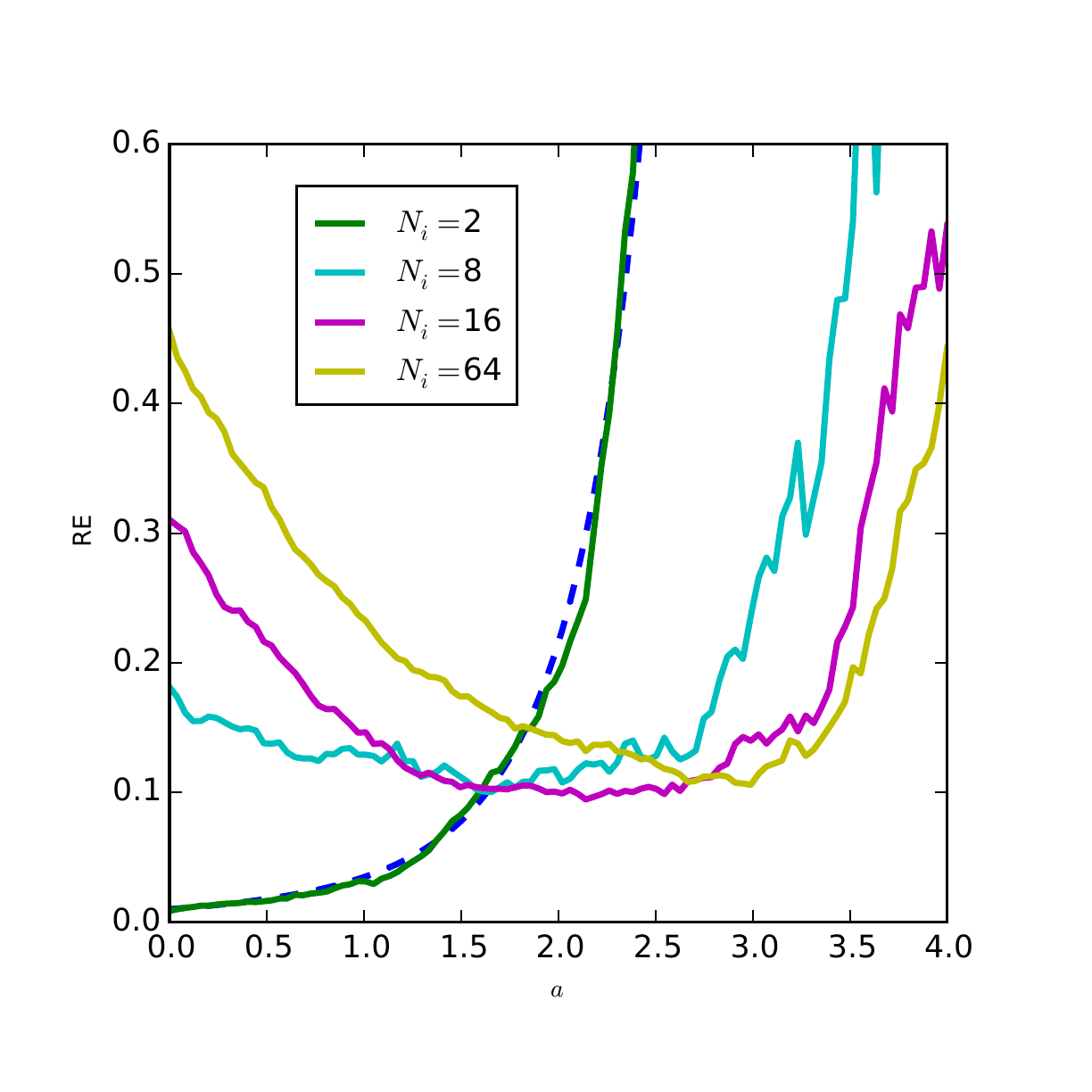}}\subfloat[]{\centering{}\includegraphics[width=0.45\textwidth]{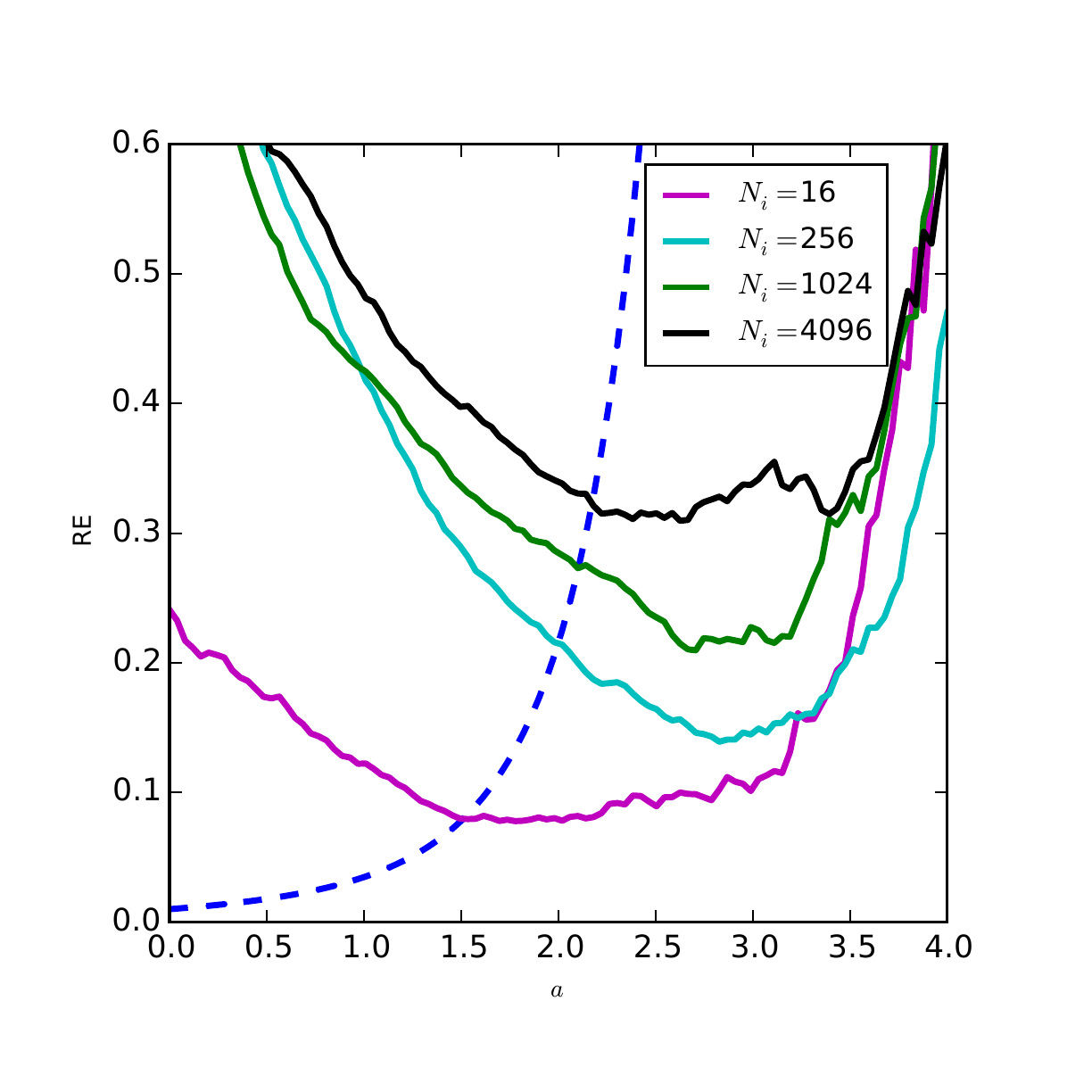}}

\caption{The estimator variance $RE=\sqrt{\frac{1}{K}\sum_{i=1}^{K}(\breve{\gamma}_{A,i}-\gamma_{A})^{2}}/\gamma_{A}$
where $A=[a,+\infty)$ with different numbers of selection steps,
with $M=1,000$ particles each. The estimator variance for brute force
Monte Carlo is shown as the blue dashed line. The weight function
is $\exp(C\Delta x)$ with $C=4$.\label{fig:numselections}}
\end{figure}

This result can be interpreted as follows. Selections shouldn't be performed too frequently, as cloning increases correlations between the particles
and therefore reduces the effective number of independent particles, increasing
the estimator variance. If not performing selections frequently enough
however, the particle distribution relaxes to the unbiased particle
measure, leading to the poor brute force Monte-Carlo variance. This
can be seen in Figure \ref{fig:numselections}: for low thresholds $a$,
importance sampling is useless and estimations with a small number of selections have
the lowest estimator relative error $RE$. Due to the large time between
selections, the particles have relaxed to the particle measure of
brute force Monte Carlo simulations and therefore have a similar estimator
variance. For higher thresholds, for instance for $a>1.7$, it becomes
advantageous to kill a larger number of particles to obtain
a more skewed final particle distribution, in order to lower the variance.
For the threshold value $a=2$ the optimal number of interactions
among the values in the figure is $N=16$. For higher thresholds there
is a small reduction in error by increasing the number of selections,
although increasing the number of selections much further beyond $N=64$
results in an overall increase of error.

Figure \ref{fig:numselections} also illustrates the large estimator
variance improvement for the genealogical particle analysis algorithm
compared with Monte-Carlo sampling, as soon as $a\geq2$.\\

Besides the number of selections, there also seems to be little theoretical
understanding of the optimal timing for selections. One strategy to
selection timing is to calculate on-the-fly a criterion on the distribution
of particle weights (such as the squared coefficient of variation
or entropy) and only perform selection if a fixed threshold is exceeded.
The convergence of such adaptive selection strategies is discussed
in \cite{del_moral_adaptive_2012}.

\subsubsection{Estimating a range of over-threshold probabilities\label{sub:Estimating-the-tail}}

In the following the weight function $W(x,y)=\exp(C(x-y))$ is used. From
the point of view of the estimator variance $RE$ (\ref{eq:Estimated_RE}),
to each value of the threshold $a$ corresponds an optimal value of
$C$, denoted $C^{*}(a)$, or equivalently for each value of $C$
the estimator variance has a minimum for a given value of $a$, denoted
$a^{*}(C)$. For instance Figure \ref{fig:numselections} shows that
the value $C=4$ is optimal for $a\simeq2.5=a^{*}$. In simple cases, we
expect $C^{*}(a)$ to increase monotonically with $a$.

\begin{figure}
\subfloat[]{\begin{centering}
\includegraphics[width=0.45\textwidth]{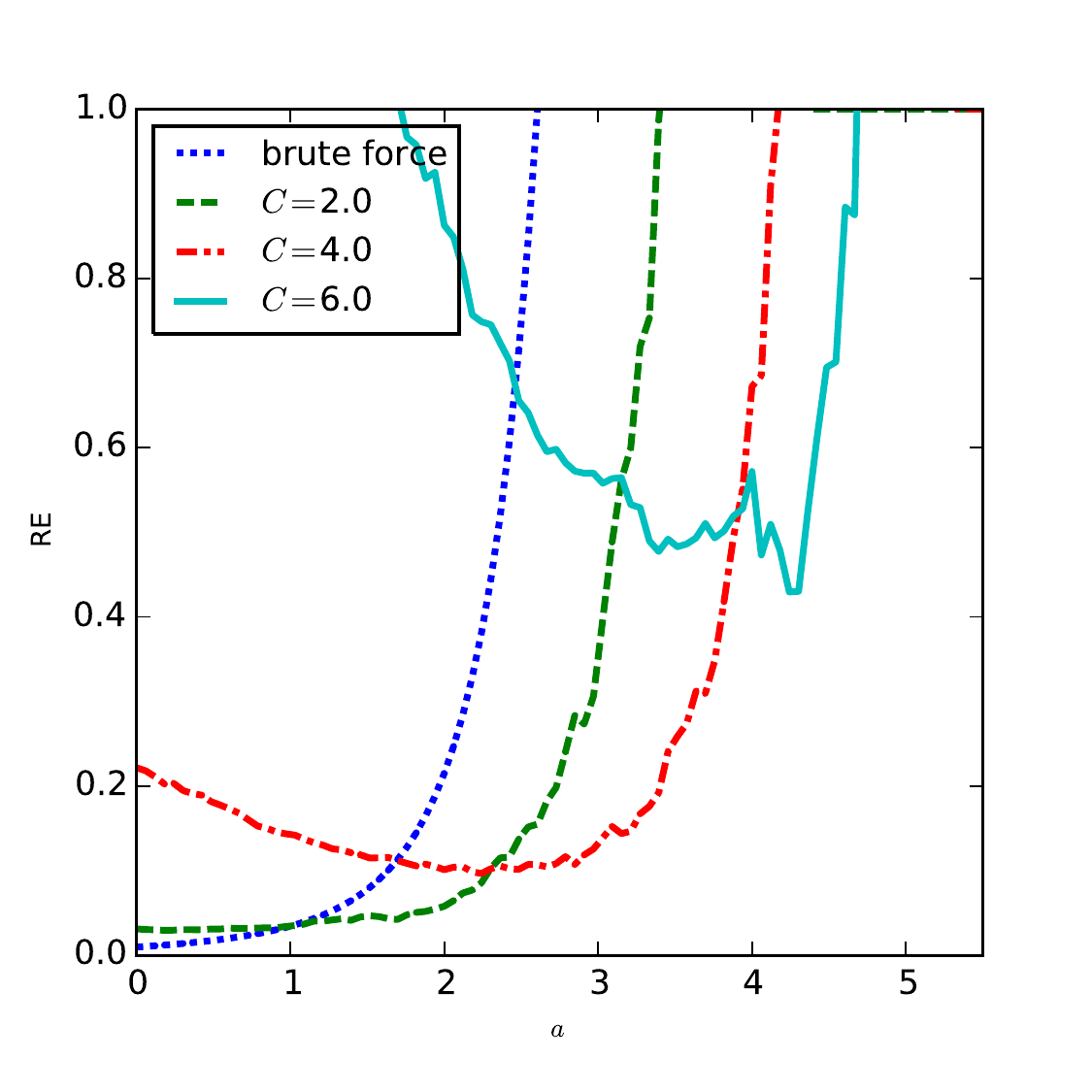}
\par\end{centering}

}\subfloat[]{\begin{centering}
\includegraphics[width=0.45\textwidth]{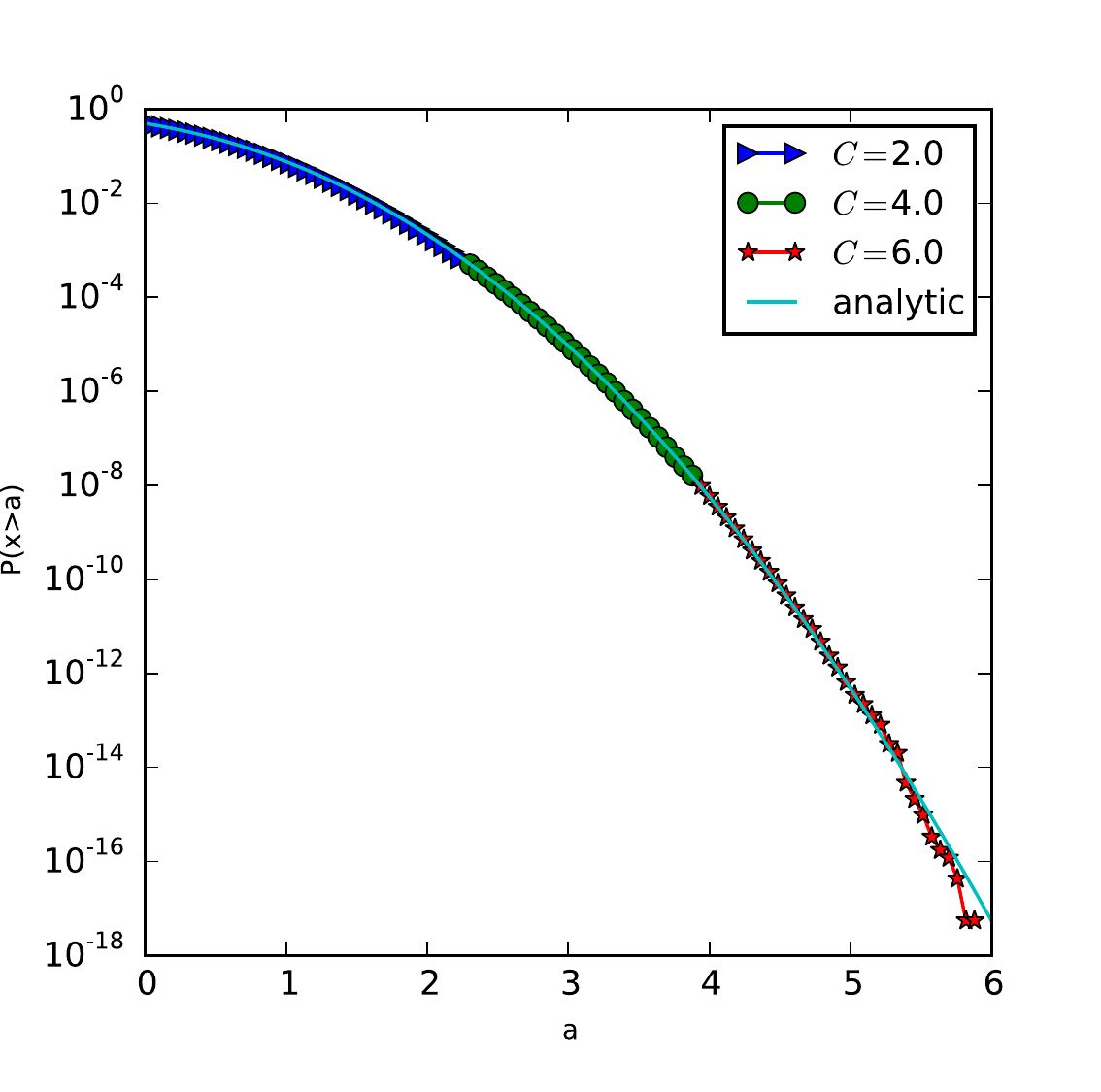}
\par\end{centering}

}

\caption{(a) The estimator variance $RE$ for different weight factors $C$
for a range of thresholds. The brute force error is computed as $\sqrt{\gamma_{A}-\gamma_{A}^{2}}/(\sqrt{M}\gamma_{A})$ where $A = [a,+\infty)$.
(b) The estimated over-threshold probability $P(a)$ compared to the
analytic result. For each value of the threshold $a$, the estimate
corresponding to the value of $C$ with lowest estimator variance
is chosen. \label{fig:tail-estimation}}
\end{figure}

There is an optimal value of $a$ for each value of $C$, however
the estimate is good for a range of thresholds around this optimum.
When instead of a particular over-threshold probability one is interested
in the tail of the complete distribution probability, one can perform
a number of genealogical particle analysis simulations each with different
value of $C$, and select for each threshold the value corresponding
to the lowest estimator variance $RE$. Figure \ref{fig:tail-estimation}
illustrates how the tail of $P(a)=P(x(t)\geq a|x(0)=0)$ can be estimated
this way, for $x$ the Ornstein-Uhlenbeck process. For large values of
the threshold (above $a\approx4.6$) all estimates have a high error
and the highest value of $C$ is chosen by default. As can be seen
on this figure there is very good agreement with the theoretical value up to probabilities as low as $10^{-10}$. Using this strategy,
we can accurately estimate the tail of the over-threshold distribution
down to probabilities as small as $10^{-10}$, with relative error lower
than one.

\subsubsection{Selections along the fluctuation paths\label{sub:Selections-along-the}}

We have discussed in Section \ref{sub:Time-dependent-weighting} that
a time-dependent weighting function can be used. In this way the particle
distribution can be weighted with different exponential factors $C(t_k)$ all along the path,
but still lead to the same exponentially tilted final particle distribution.
Furthermore, in Section \ref{sub:Fluctuation-paths} we have discussed
how for small noises, most of the paths leading to a rare event will
concentrate around fluctuation paths that minimize the action functional.
The aim of this section is to demonstrate the interest of using fluctuation
paths to construct time-dependent weighting functions in order
to increase the efficiency of the genealogical particle analysis algorithm.

Since the Ornstein-Uhlenbeck process is linear, taking limits
of higher thresholds is equivalent to taking a weak-noise limit through
a rescaling of the $x$ coordinate. Hence, for fixed noise intensity
$\sigma$, paths starting at $x_{0}=0$ conditioned on reaching the
final threshold $a$ will concentrate around the fluctuation paths
in the limit $a\rightarrow\infty$. The action (\ref{eq:Action}) for
the Ornstein-Uhlenbeck process (\ref{eq:ornstein-uhlenbeck}) is given
by $I\left[X\right]=\int_{0}^{T}d\tau(\dot{X}+\lambda X)^{2}$. Taking
as boundary conditions $X(0)=0$ and $X(T)=a$ the fluctuation paths
are easily computed to be $X_{f}(t)=a\frac{\sinh\lambda t}{\sinh\lambda T}.$

By using the potential function $W(t,x,y)=\exp(C(t_{k})x-C(t_{k-1})y)$
with a weight parameter $C(t_{k})$ dependent on the selection
time $t_{k}$, we can control $\tilde{\mu}(t_{k})$, the mean particle
position at $t_{k}$, by fixing $C(t_{k})$. The expected particle distribution
for the Ornstein-Uhlenbeck process tilted with this weighting function
after the selection at $t_{k}$ is 
\[
\exp\left(C(t_{k})x\right)\mathcal{N}_{0,v(t_{k})}(x)/\int\mbox{d}x\,\exp\left(C(t_{k})x\right)\mathcal{N}_{0,v(t_{k})}(x)
\]
 as discussed in Section \ref{sub:Time-dependent-weighting}. The
corresponding expected mean particle position is therefore $\tilde{\mu}(t_{k})=C(t_{k})v(t_{k})$
where $v(t)=(1-\exp(-2t))/2$ is the variance of the Ornstein-Uhlenbeck
process at time $t$ (the solution of Eq. \ref{eq:v} with $v(0)=0$).
Choosing $C(t_{k})=X_{f}(t_{k})/v(t_{k})$, $\tilde{\mu}(t)$ follows
$X_{f}(t)$ and the algorithm particle distribution closely follows
the fluctuation path leading to the threshold $a$.

\begin{figure}
\begin{centering}
\includegraphics[width=0.8\textwidth]{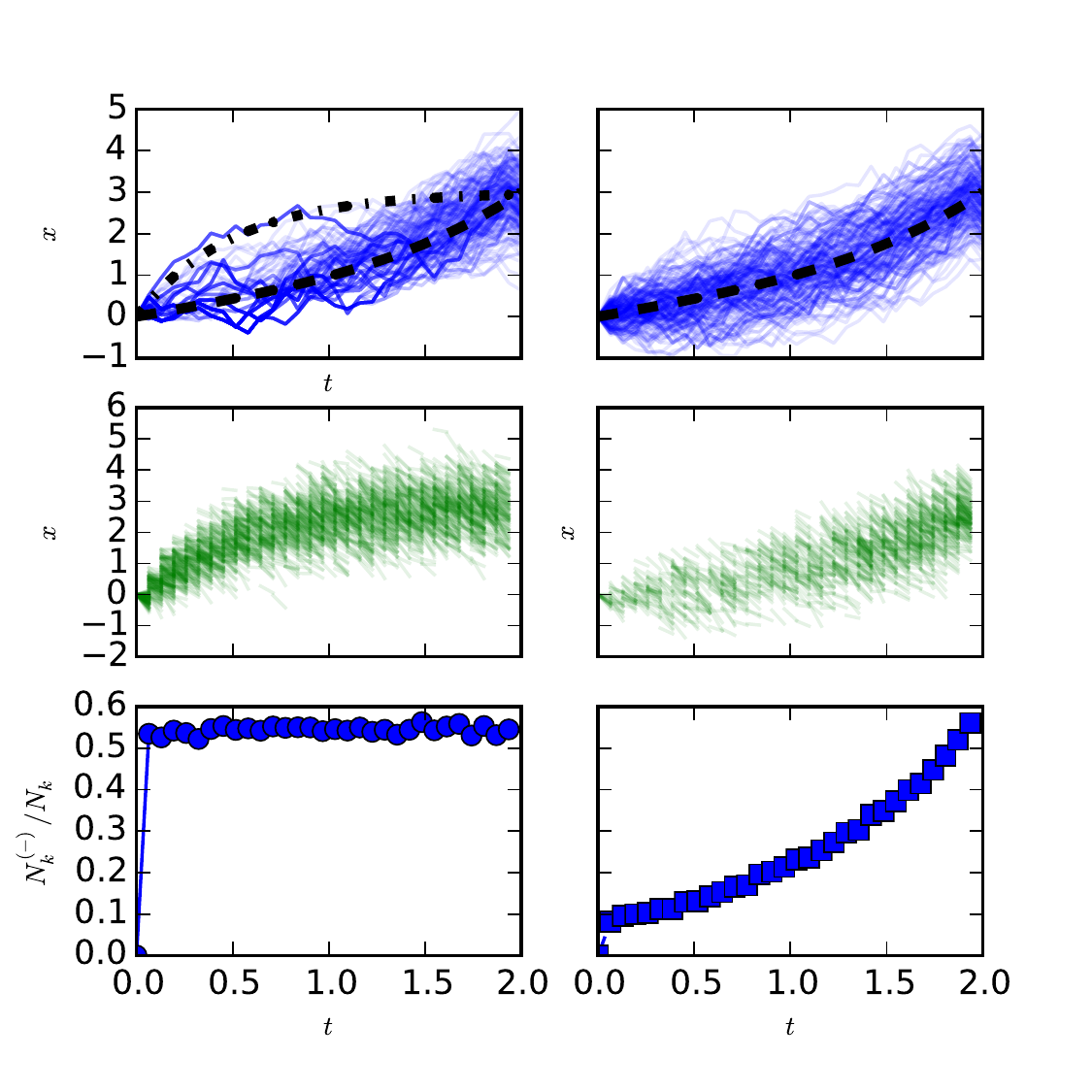}
\par\end{centering}

\caption{Ancestral paths (top), the final portion of killed paths, plotted
only between $t_{k-1}$ and $t_{k}$ if the path is killed at $t_{k}$
(middle) and the fraction of the number of particles killed $N_{k}^{(-)}$
to the total number of particles $N_{k}$ (bottom) for genealogical
particle analysis algorithms with either exponential weighting with
$C=6.0$ (left) or weighting based on the fluctuation path (right)
for the fluctuation path ending at $a=3.0$ at the final time $T=2$.
The dashed black lines in the top plots show the fluctuation path.
The dash-dotted line in the top left plot shows the mean of the target
particle distributions after selection (equals $Cv(t)$). The average
number of particles for both simulations is $M=10^{4}$ and the number
of selections steps is $32$. For graphical purposes a randomly selected
sample of 2\% of the ancestral and killed paths are shown in the first
two rows \label{fig:ancestral-paths}}
\end{figure}

\begin{figure}
\begin{centering}
\includegraphics[width=0.75\textwidth]{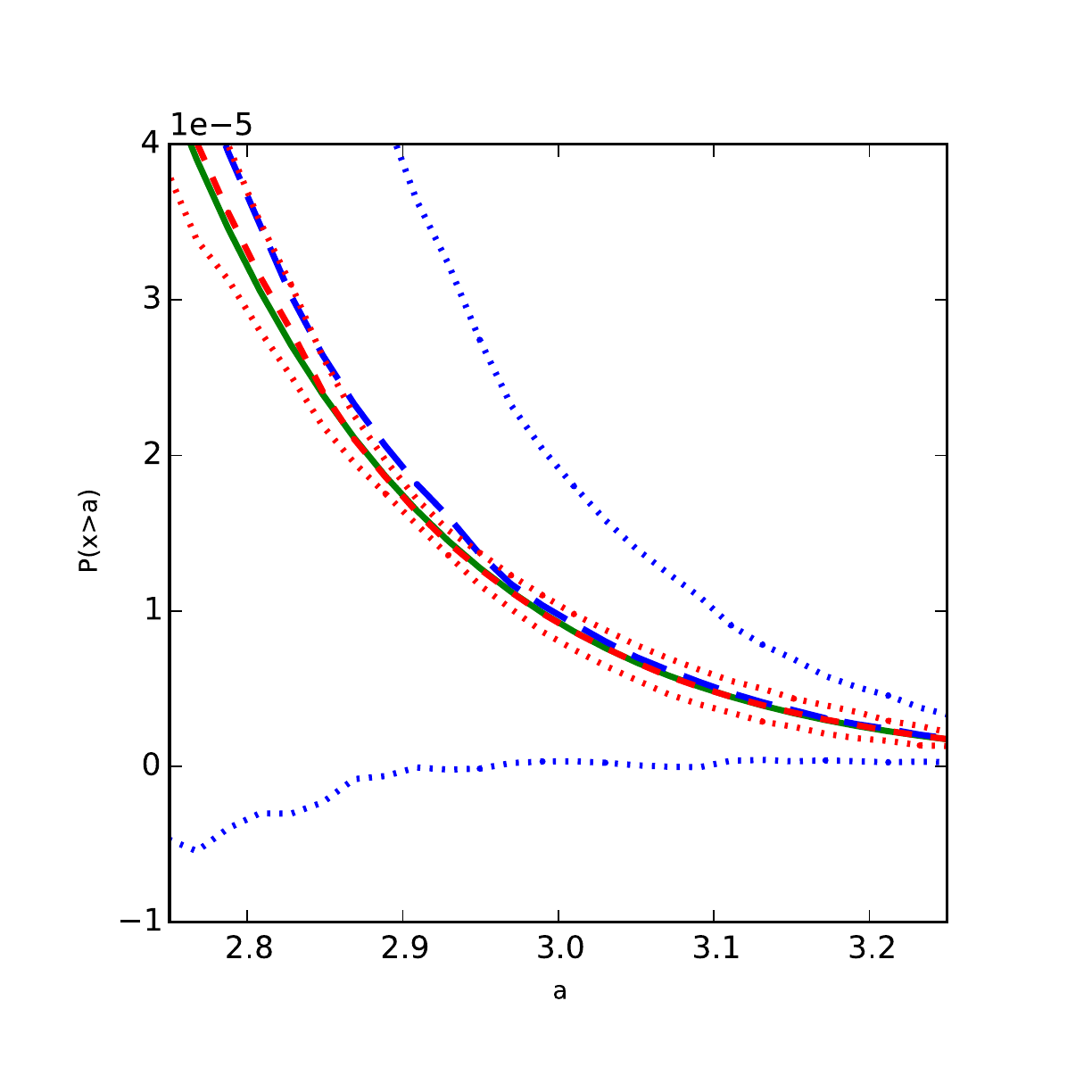}
\par\end{centering}

\caption{The over-threshold probability $P(x>a)$ as estimated by the genealogical
particle analysis algorithm either with an exponential weight (blue
long-dashed line) or a weight based on fluctuation paths (red medium-dashed
line). The two short-dashed lines, at equal distance from the estimated
averages, correspond to a 2 standard deviation interval of the estimator.
The full line is the analytic result. Both implementations use the
same number of particles \textbf{$N=1e4$} and $32$ selections
steps and both have roughly the same computational cost.\label{fig:instanton-variance}}
\end{figure}

Figure \ref{fig:ancestral-paths} shows the effect of using a weight
function based on a fluctuation path versus an exponential weight
function. The bottom two plots show how using the fluctuation path
significantly decreases the fraction of particles that are killed
during the selection steps ($N_{k}^{(-)}$) to the number of particles
at that time step ($N_{k}$). This is also illustrated in the plots in the top and middle rows. The top plots show the paths from the initial state for all surviving particles at the final time. Paths that have been killed during the process
are not shown. We call these paths the ancestral paths. As can be
seen on the top left plot, only few trajectories from the initial stage
of the algorithm are ancestors of the final positions. This is
not the case for the top right plot. The algorithm using a weight
based on a fluctuation path has a much larger number of ancestors.
This richer ancestral tree results in a decreased estimator error
for the over-threshold probabilities, as is demonstrated in
Figure \ref{fig:instanton-variance}.

Note that for both the exponential weighting function and the weighting
based on the fluctuation path, the paths reaching the threshold follow
the fluctuation path. Other paths reaching the threshold are so rare
that few of them are generated, even if they are more likely to survive
selection in the case of exponential weighting. Note that the killed
paths, partially shown in the middle row of Fig. \ref{fig:ancestral-paths}, tend to have a negative change in position before being killed.
The higher target path in the exponential makes for a higher average
dissipative force $-x$ on the particles, leading to a large discrepancy
between the actual particle distribution and the target distribution
at selection times.

\section{Genealogical particle analysis algorithm for a deterministic dynamical
system: the Lorenz '96 model\label{sec:Interacting-particle-algorithm}}

The Lorenz '96 model is a deterministic dynamical system that is often
used as a toy model in the meteorology community. It was proposed
by Lorenz as part of a study on error growth and predictability for
chaotic dynamical systems \cite{lorenz_predictability:_1996,lorenz_designing_2005}.

A crucial difference between the famous Lorenz '63 model and the less
well-known Lorenz '96 model is that the latter has a large number
of degrees of freedom. Indeed macroscopic variables of deterministic
systems with a large number of degrees of freedom often behave qualitatively
similar to solutions of stochastic differential equations with much
less degrees of freedom. Such results can be proven for some specific
types of models (feauturing separation of time scale, independence, ...).

It is believed however that similar results remain true for a wide
range of models and observables even though mathematical proofs are
out of reach. If this conjecture is correct, then the sampling of
rare events through genealogical particle analysis algorithms should
be applicable to macroscopic variables of deterministic systems with a large
number of degrees of freedom. In this section, we demonstrate empirically,
through numerical simulation, that genealogical particle analysis
algorithms can indeed efficiently sample the tail of the energy
distribution for the Lorenz '96 model.

\subsection{Description of the model\label{sub:Description-of-the}}

The Lorenz '96 model consists of $L$ variables $x_{i}$ on a ring
$i\in\{0,..,L-1\}$, with dynamics 

\[
\dot{x}_{i}=x_{i-1}(x_{i+1}-x_{i-2})+R-x_{i}
\]
where indices $i$ are in $\mathbb{Z}_{L}$, i.e. the index $i$ is
identified with $i\bmod L$ if $i\notin\{0,..,L-1\}$. The non-linear part of the dynamics
$x_{i-1}(x_{i+1}-x_{i-2})$ conserves the energy $E(x)=\frac{1}{2L}\sum_{i=1}^{L}x_{i}^{2}$,
while $R$ is a forcing and $-x_{i}$ a linear dissipation. The dynamics
is chaotic for $R\geq8$ \cite{lorenz_predictability:_1996,lorenz_designing_2005}.
We will estimate the probability of reaching a certain energy threshold
after a time $t$, starting from the zero vector \textbf{$x_{0,i}=0\,\forall i$}. \referee{A small perturbation $\epsilon\mathcal{\vec{N}}_{0,1}$ is added to the particle configuration to make the trajectories diverge. For large enough times $T$, the system will therefore relax to its physical invariant measure and it makes sense to determine probabilities of exceedances of macroscopic observables $\gamma_{E_t}:=P(E(x(T))> E_t)$.}
Throughout the article we will use a number of variables $L=32$ and
a forcing $R=2^{8}=256$.

\begin{figure}
\begin{centering}
\includegraphics[width=0.5\textwidth]{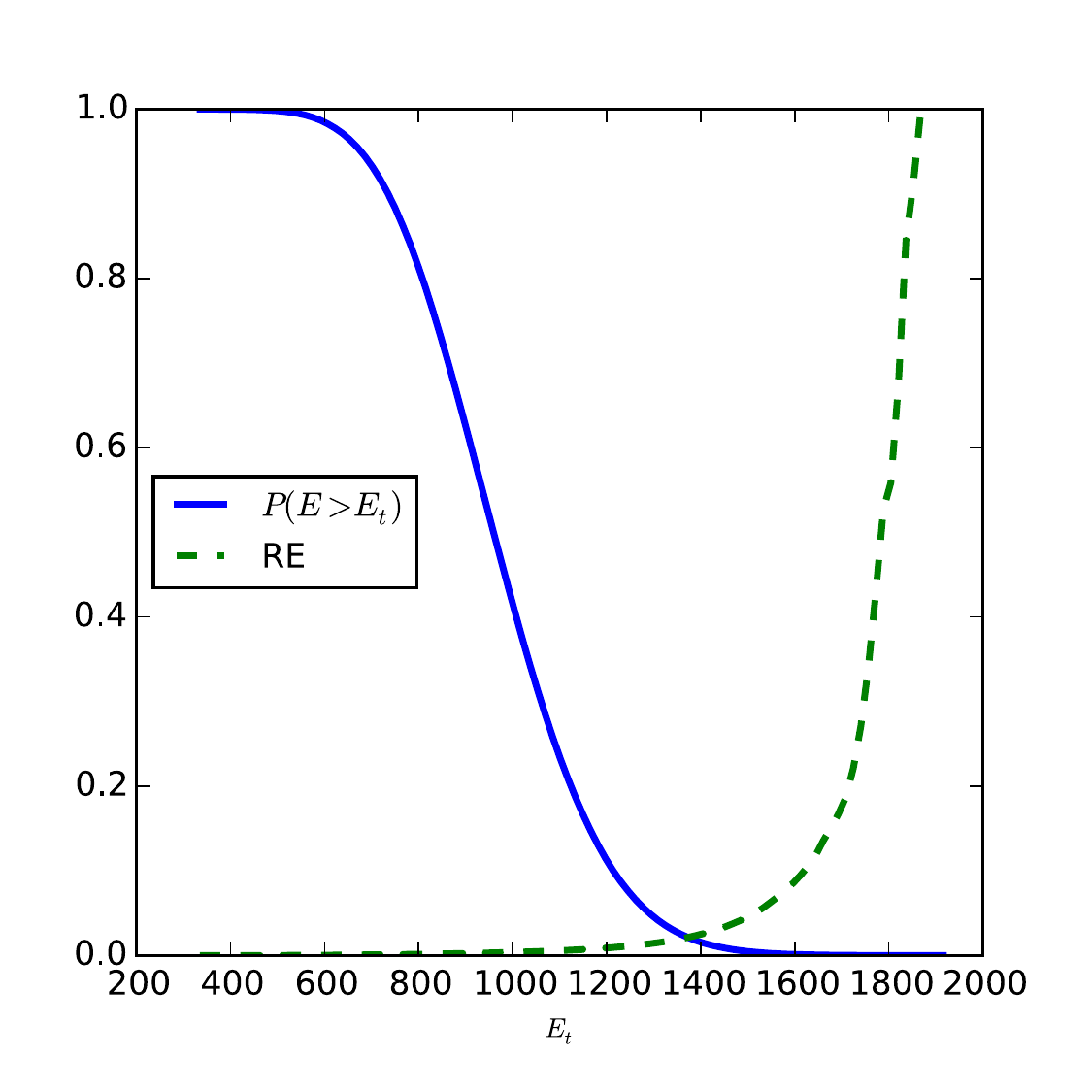}
\par\end{centering}

\caption{Over-threshold probability $\gamma_{E_t} = P(E(t)>E_t)$ estimated from a brute force simulation
and estimator variance $RE=\sqrt{\gamma_{E_t}-\gamma_{E_t}^{2}}/(\sqrt{M}\gamma_{E_t})$
of the energy $E$ of the Lorenz '96 system with $R=2^{8}$ and $M=10^{5}$. }
\label{fig:energy_hist}
\end{figure}

Figure \ref{fig:energy_hist} shows a plot of the over-threshold probabilities
of the energy of the Lorenz '96 system, estimated through a brute
force Monte-Carlo simulation from randomly perturbed initial conditions. Given that we have finite computer resources
at our disposal, assume we can generate at most $M=10^{5}$ independent
measurements of the energy. If the maximal relative error that we
are willing to tolerate is for example $0.5$ then since $RE=\sqrt{\gamma_{E_t}-\gamma_{E_t}^{2}}/(\sqrt{M}\gamma_{E_t})\approx1/(\sqrt{M\gamma_{E_t}})$
the lowest probability that we can estimate is approximately $\gamma_{E_t}=1/M(RE)^{2}=4.10^{-5}$.
From Figure \ref{fig:energy_hist} we can deduce that the corresponding
highest energy threshold obtainable lies around an energy threshold
$E_{t}=1785$. Beyond this threshold the use of rare event algorithms
becomes necessary.

\subsection{Algorithm implementation}

We use the following settings for the genealogical particle analysis
simulation. The initial condition is set to \referee{$x_{i}=\epsilon\mathcal{\vec{N}}_{0,1}$}. The total
integration time per realization is \textbf{$T=1,27$}. This time interval corresponds to roughly $5$ times the decorrelation time of the energy observable.
 The standard
deviation of the estimator $\sqrt{\frac{1}{K}\sum_{i=1}^{K}(\breve{\gamma}_{A,i}-\gamma_{A})^{2}}$
is estimated from $K=10$ independent runs of the algorithm and the
truth $\gamma_{A}$ is taken from a long brute force Monte Carlo simulation.
The number of interaction is set to $64$.

\subsubsection{Weight function\label{sub:Weight-function}}

For simplicity, we have employed an exponential weight function $W=\exp(C\Delta E)$
where $\Delta E$ is the change in energy between two interactions.
This choice doesn't require any a priori knowledge of the dynamics
and is easy to implement. This weight function is
not optimal, but, as we will show, it already gives good results.

For the value of the forcing parameter $R=2^{8}$ the distribution
of the energy values is roughly Gaussian. One can
therefore estimate the mean $\mu_{E}$ and the variance $\sigma_{E}^{2}$
from a brute force Monte Carlo simulation and use these values along
with the reasoning of Section \ref{sub:Skewing-a-normal} to determine
an appropriate value of the exponential weighting factor $C$ in the
weighting function $W=\exp(C\Delta E)$. One can then choose a value
$C=\Delta\mu_{E}/\sigma_{E}^{2}$ where $\Delta\mu_{E}$ is the desired
change of the mean energy of the final particle distribution.

The values of $C$ in the weighting function $W=\exp(C\Delta E)$
for the calculations presented in this section are taken as $C_{r}=r/(2\sigma_{E})$
with $r\in\{1,2,3,,4\}$ and $\sigma_{E}$ being the standard deviation
of the energy so as to increase the mean energy $\Delta\mu_{E}$
by steps of size $\sigma_{E}/2$.

\subsubsection{Noise perturbation}

For deterministic dynamical systems, in order for two trajectories
to have different dynamics after selection, a small perturbation can
be added. This can be achieved by adding for example a weak Brownian
perturbation at all times, or by adding a small instantaneous perturbation
to offspring at the selection times. The former approach provides
a simpler mathematical framework. Indeed the study of the noise effect
would amount to the study of the stochastic differential equation
properties in the weak noise limit, independent of the genealogical
particle analysis algorithm. By contrast the latter approach intertwines
the random perturbation with the genealogical particle analysis algorithm
effects and is therefore more complicated to analyze. The latter approach,
however, has the practical advantage of being computationally simpler.
In this study, as we will proceed purely empirically, we have opted
for the latter approach. The clones are perturbed by $\epsilon\mathcal{\vec{N}}_{0,1}$
where $\mathcal{\vec{N}}_{0,1}$ is a standard $L$-dimensional Gaussian
random variable, i.e. the noise acts independently on all of the variables.

\begin{figure}
\subfloat[]{\includegraphics[width=0.45\textwidth]{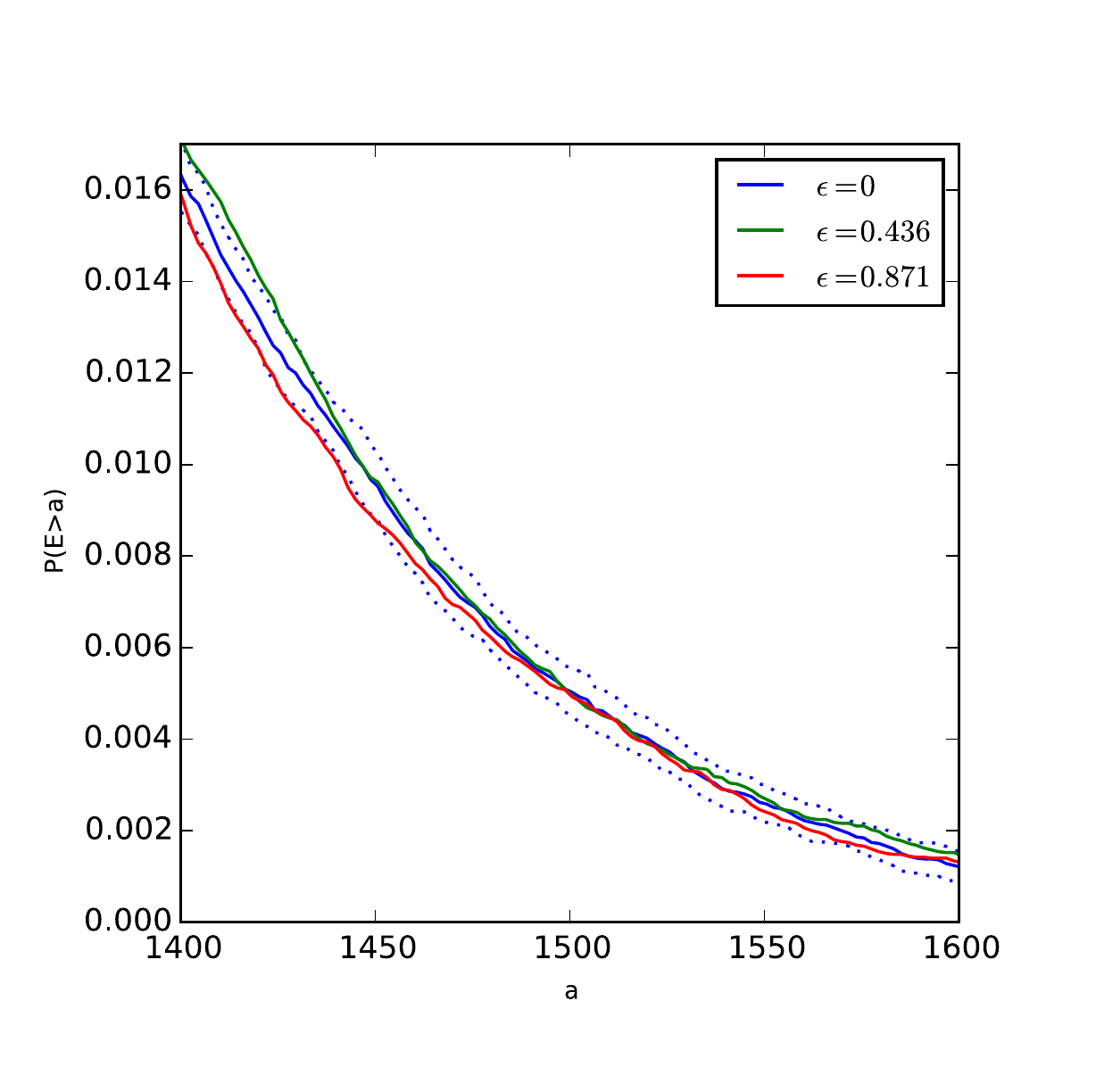}\label{fig:epsilons-1}

}\subfloat[]{\includegraphics[width=0.45\textwidth]{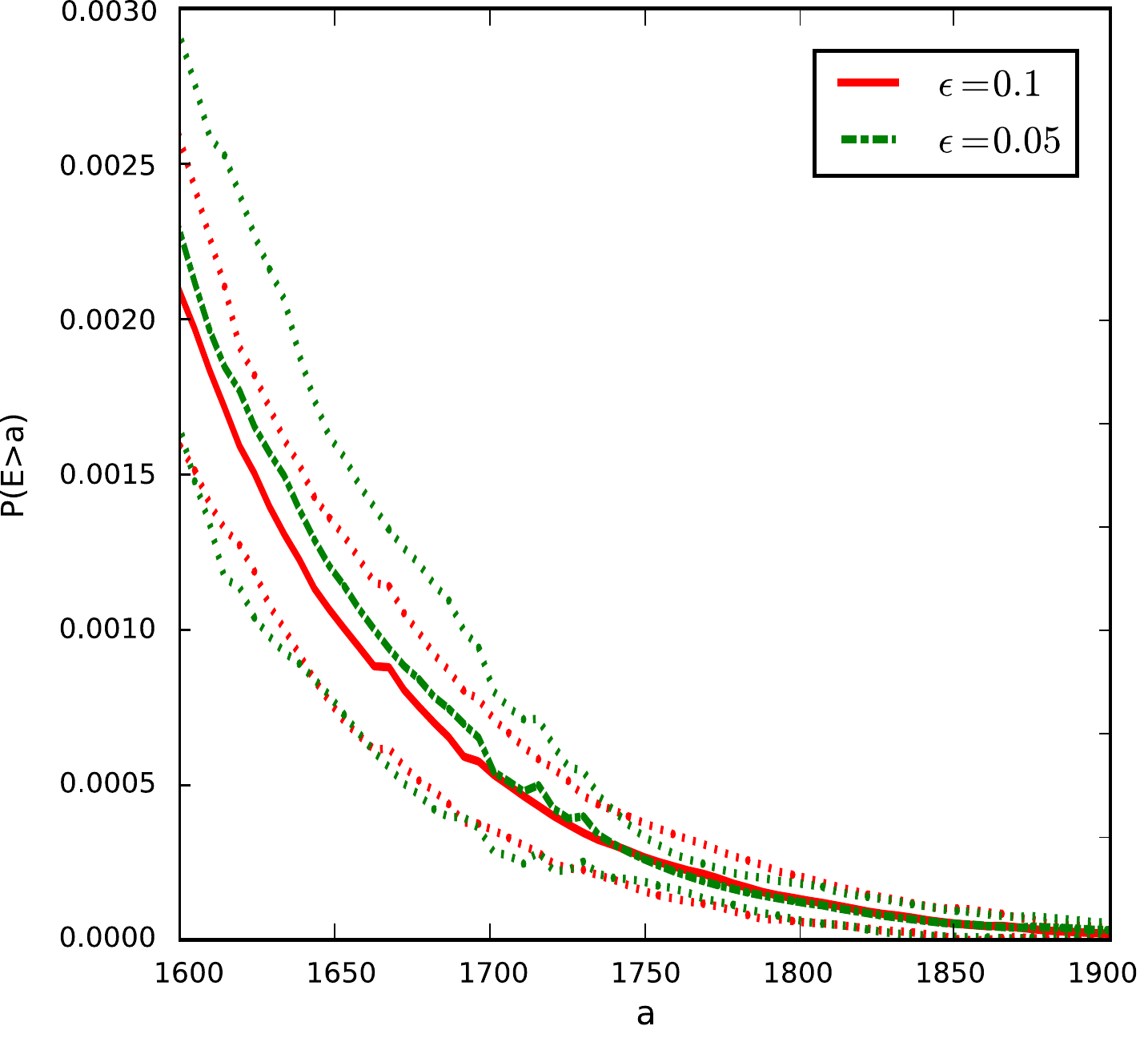}\label{fig:epsilons}

}

\caption{(a) The over-threshold probability $P(E>a)$ of the energy $E$ of
the Lorenz '96 system with perturbations of varying strengths $\epsilon$
at times $t_{k}$, without performing killing and cloning, with 10 000 independent realisations. The dotted line shows the estimated $2 \sigma$ interval for 
over-threshold probability of the energy of the Lorenz '96 system,
as estimated from two realizations of the genealogical particle analysis
algorithm with different perturbing noise strength $\epsilon$ upon
cloning.}
\end{figure}

The small noise perturbation invariably adds an error to the estimates
of the tail probabilities. To obtain a rough upper bound on the strength
of the perturbation that can be added without significantly perturbing
the tail, we first perform a brute force simulation with the added
noise for different noise strengths and verify that the tail probabilities
do not change significantly compared to the sampling error of the
brute force calculation. A set of independent realizations is performed
like in the brute force Monte Carlo approach, the only difference
being that at the selection times $t_{k}$ the same noise perturbations
is added as in the genealogical particle analysis simulation. No selection
is performed however in these runs. This way we can estimate the effect
of the noise on the final time particle distribution. Figures \ref{fig:epsilons-1} and \ref{fig:epsilons} show that below a perturbation strength
of $\epsilon=0.87$ and for thresholds higher than $a=1600$, the
noise does not have a significant effect on the over-threshold probabilities.
More complex schemes of noise perturbation could be implemented to
assure that the perturbed trajectory remains close to the attractor,
for example by storing a configuration at a time point before $t_{k}$,
adding a small perturbation to it and evolve it up to $t_{k}$ to have
the perturbation relax towards the attractor.

Furthermore, after performing the genealogical particle analysis algorithm,
we check that the perturbing noise intensity $\epsilon$ is small
enough by decreasing $\epsilon$ and checking that the estimates of
the over-threshold tail statistics are consistent. Figure \ref{fig:epsilons}
shows that for $\epsilon=0.1$ the results remain stable upon halving
the noise intensity.

\subsubsection{Estimating the tail of the energy distribution}

\begin{figure}
\subfloat[]{\includegraphics[width=0.45\textwidth]{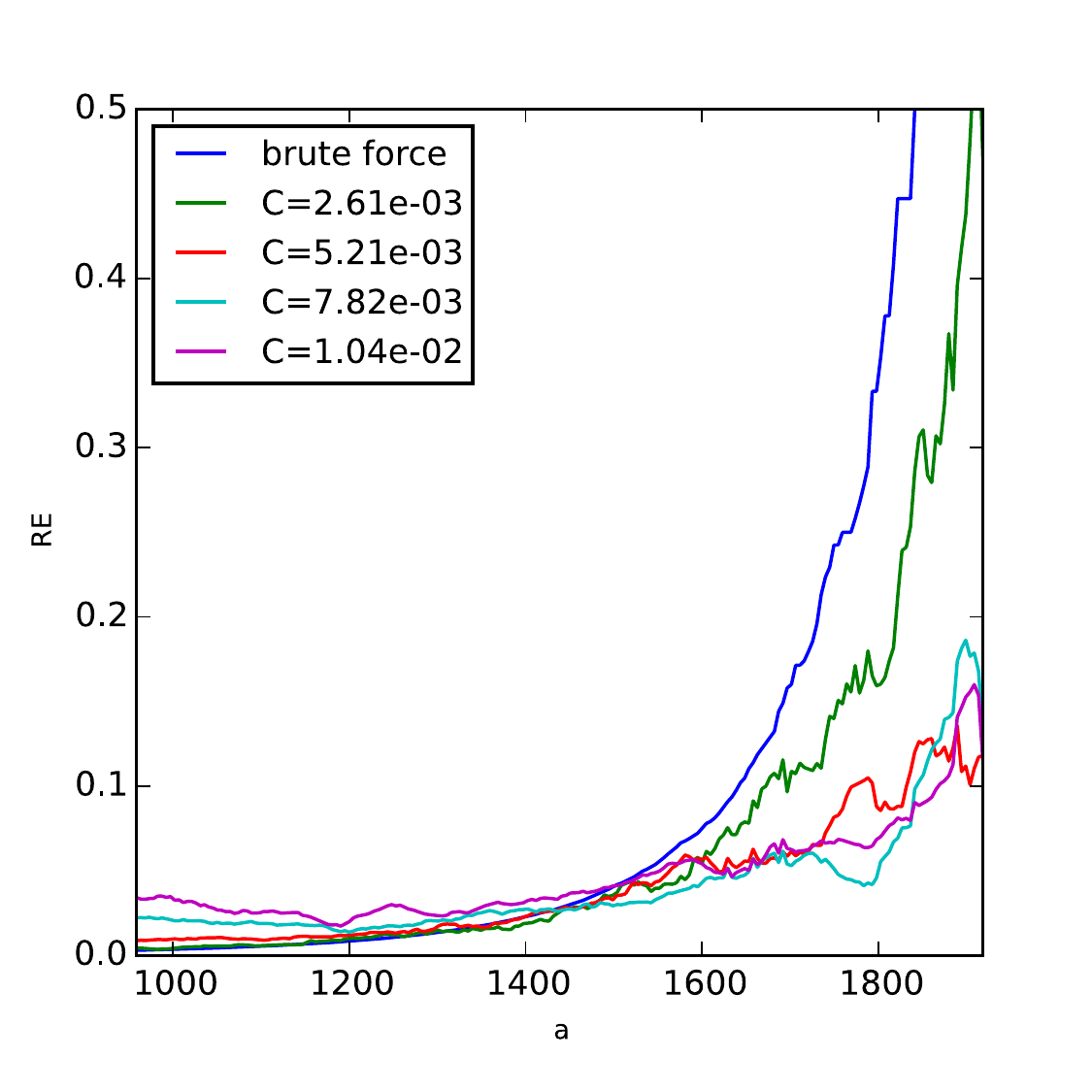}

}\subfloat[]{\includegraphics[width=0.45\textwidth]{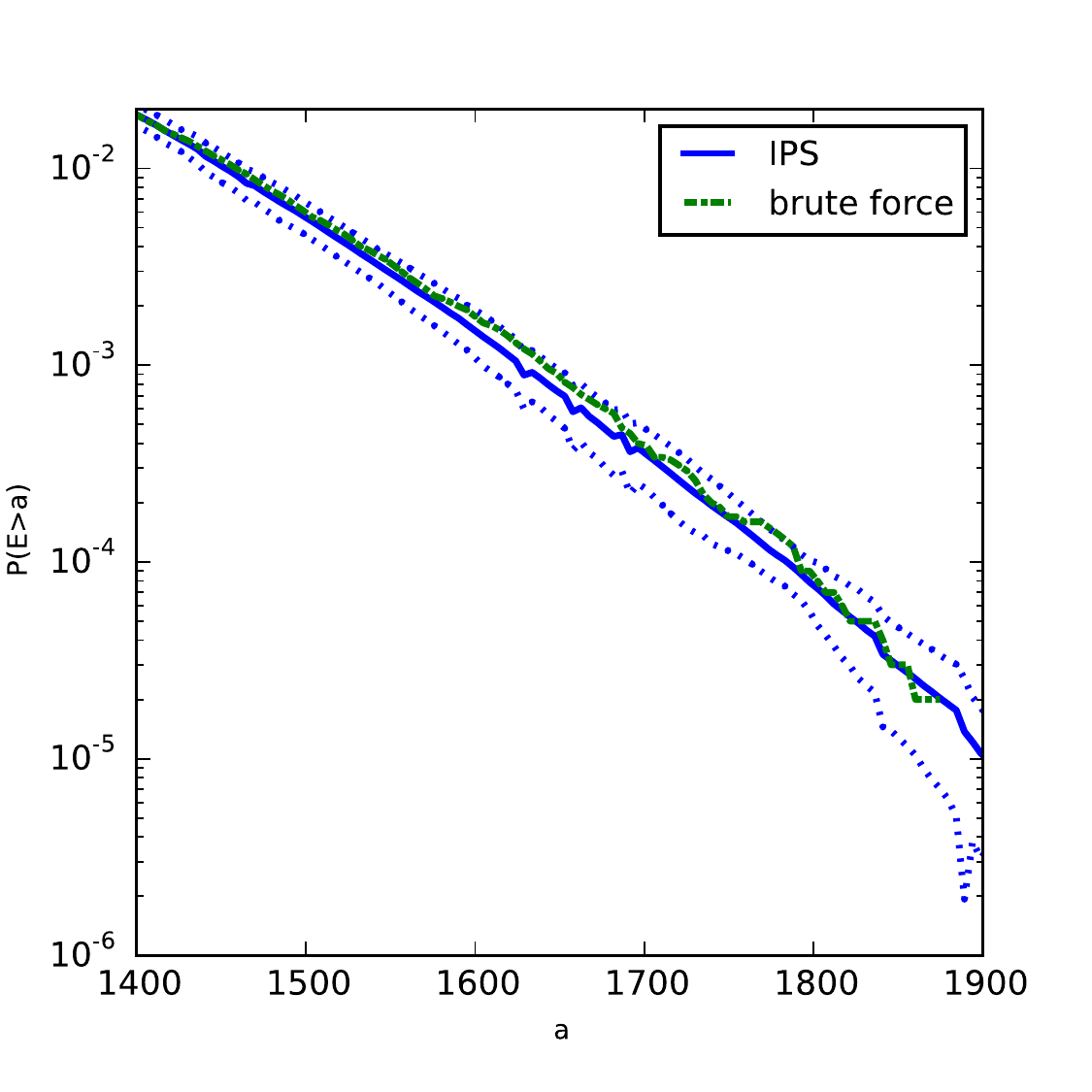}

}

\caption{(a) The empirical relative error $RE$ for different weight factors
$C$, for a range of energy thresholds $a$, for the Lorenz '96 model.
The number of particles is $M=1,000$. The brute force Monte Carlo
relative error (in blue) is estimated with the same number of realizations
$M$ as $\sqrt{\hat{\gamma}_{a}-\hat{\gamma}_{a}^{2}}/(\hat{\gamma}_{a}\sqrt{M})$
(b) The over-threshold probability tail as estimated from the genealogical
particle analysis algorithm compared to a brute force reconstruction.
The number of particles used is $1,000$ for the genealogical particle
analysis simulation and $10,000$ for the brute force simulation.}
\label{fig:l96_tail_estimation}

\end{figure}

We use for the Lorenz '96 model the procedure described in Section
\ref{sub:Estimating-the-tail}: we increase the values of the weight
parameter $C$ and use for each threshold value the best estimate
from the point of view of the empirical estimator variance. The result
is shown in Figure \ref{fig:l96_tail_estimation}. As there is no
analytic expression for the energy distribution tail of the Lorenz
'96 system, we use a long brute force Monte Carlo estimation
as comparison. The estimator variance markedly decreases when using
the genealogical particle analysis algorithm. When constructing the
over-threshold probability, we see that the tail can be reliably reproduced
when compared to the longer brute force calculation.

The improvements in efficiency from using a rare event simulation
scheme can be quantitatively estimated from Figure \ref{fig:l96_tail_estimation}.
The plot of the empirical relative error shows how for a threshold
around $a=1800$ a brute force Monte Carlo calculation yields a relative
error of $0.5$, whereas the genealogical particle analysis simulation
yields a relative error of approximately $0.05$. A reduction in relative
error by a factor $10$ is achieved. Since the brute force Monte Carlo
error scales as $1/\sqrt{M}$, a similar reduction by a raw increase
of processing power would require $M$ to increase by a factor of
$100$. For higher thresholds and with more fine tuning of the selection
process, a much larger reduction is likely to be achievable.

\section{Conclusion\label{sec:Conclusion}}

In this paper we have addressed the use of rare event computation
techniques to estimate small over-threshold probabilities of observables
in deterministic dynamical systems. We have demonstrated that the
genealogical particle analysis algorithms can be successfully applied
to a toy model of atmospheric dynamics, the Lorenz '96 model as presented
in Section \ref{sub:Description-of-the}. We have furthermore used
the Ornstein-Uhlenbeck system to illustrate a number of implementation
issues.

The example of the Ornstein-Uhlenbeck has illustrated the importance
of the choice of the objective function for the performance of the
genealogical particle analysis algorithm estimator. We have shown
how a time-dependent objective function based on the fluctuation path
to a high threshold can greatly improve the performance of the estimator
compared to a fixed-in-time objective function. Furthermore we have
discussed how the number of particles and the number of selection
steps influence the performance of the estimator. 

For the deterministic chaotic system a complication arises in that
a stochastic perturbation needs to be added to the system to make
identical clones of one parent diverge and explore the system's path
space. We have demonstrated in this example how the estimates of the
rare event simulation are stable for small perturbations and agree
with results from brute force Monte Carlo estimations. We therefore
can have confidence in the correctness of these estimates.

For the example of a deterministic chaotic system that we have studied
we have not yet used the fluctuation path approach, since this would
require information on the dynamics to the rare event that we a priori
do not possess. This lack of knowledge can be improved by iterating
the estimation procedure, where one uses estimates of an initial brute
force simulation to estimate the fluctuation path, after which an
genealogical particle analysis simulation based on this path can be
used to estimate a higher fluctuation path, which can be used for
a next iteration of the algorithm. However, the results of the straightforward
implementation of the rare event simulation already shows significant
improvements compared to brute force estimation.


\subsection*{Acknowledgements}

The authors would like to thank Eric Vanden-Eijnden, Eric Simonnet,
Joran Rolland, Pascal Yiou, B\'ereng\`ere Dubrulle, Francesco Ragone and
Takahiro Nemoto for fruitful discussions.

JW acknowledges the support of the AXA Research Fund.

The research leading to these results has received funding from the
European Community's Seventh Framework Programme (FP7/2007-2013) under
grant agreement n° PIOF-GA-2013-626210.

\bibliographystyle{iopart-num}
\bibliography{rare_event_simulation2}

\providecommand{\newblock}{}
\begin{thebibliography}{10}
\expandafter\ifx\csname url\endcsname\relax
  \def\url#1{{\tt #1}}\fi
\expandafter\ifx\csname urlprefix\endcsname\relax\def\urlprefix{URL }\fi
\providecommand{\eprint}[2][]{\url{#2}}

\bibitem{qiu_kuroshio_2000}
Qiu B and Miao W 2000 {\em Journal of Physical Oceanography\/} {\bf 30}
  2124--2137 ISSN 0022-3670

\bibitem{schmeits_bimodal_2001}
Schmeits M~J and Dijkstra H~A 2001 {\em Journal of Physical Oceanography\/}
  {\bf 31} 3435--3456 ISSN 0022-3670

\bibitem{glatzmaier_three-dimensional_1995}
Glatzmaier G~A and Roberts P~H 1995 {\em Physics of the Earth and Planetary
  Interiors\/} {\bf 91} 63--75 ISSN 0031-9201

\bibitem{field_managing_2012}
{Intergovernmental Panel on Climate Change} 2012 {\em Managing the risks of
  extreme events and disasters to advance climate change adaption: special
  report of the {Intergovernmental} {Panel} on {Climate} {Change}\/} (New York,
  NY: Cambridge University Press) ISBN 978-1-107-02506-6

\bibitem{barriopedro_hot_2011}
Barriopedro D, Fischer E~M, Luterbacher J, Trigo R~M and Garc\'ia-Herrera R
  2011 {\em Science\/} {\bf 332} 220--224 ISSN 0036-8075, 1095-9203

\bibitem{wetter_underestimated_2013}
Wetter O and Pfister C 2013 {\em Clim. Past\/} {\bf 9} 41--56 ISSN 1814-9332

\bibitem{de_haan_extreme_2006}
de~Haan L 2006 {\em Extreme value theory: an introduction\/} Springer series in
  operations research (New York ; London: Springer) ISBN 0-387-23946-4

\bibitem{leadbetter_extremes_1983}
Leadbetter M~R 1983 {\em Extremes and related properties of random sequences
  and processes\/} Springer series in statistics (New York: Springer-Verlag)
  ISBN 0-387-90731-9

\bibitem{kharin_estimating_2005}
Kharin V~V and Zwiers F~W 2005 {\em Journal of Climate\/} {\bf 18} 1156--1173
  ISSN 0894-8755

\bibitem{kysely_probability_2002}
Kysel\'y J 2002 {\em Studia Geophysica et Geodaetica\/} {\bf 46} 93--112 ISSN
  0039-3169, 1573-1626

\bibitem{rubino_rare_2009}
Rubino G and Tuffin B (eds) 2009 {\em Rare event simulation using {Monte}
  {Carlo} methods\/} (Chichester, UK: Wiley) ISBN 978-0-470-77269-0

\bibitem{bucklew_introduction_2004}
Bucklew J~A 2004 {\em An introduction to rare event simulation\/} (New York:
  Springer) ISBN 0-387-20078-9 978-0-387-20078-1 1-4419-1893-0
  978-1-4419-1893-2

\bibitem{del_moral_mean_2013}
Del~Moral P 2013 {\em Mean field simulation for {Monte} {Carlo} integration\/}
  ISBN 978-1-4665-0417-2 1-4665-0417-X

\bibitem{moral_feynman-kac_2004}
Del~Moral P 2004 {\em Feynman-{Kac} {Formulae} {Genealogical} and {Interacting}
  {Particle} {Systems} with {Applications}\/} (New York, NY: Springer New York)
  ISBN 978-1-4684-9393-1 1-4684-9393-0

\bibitem{cerou_adaptive_2007}
C\'erou F and Guyader A 2007 {\em Stochastic Analysis and Applications\/} {\bf
  25} 417--443 ISSN 0736-2994

\bibitem{rolland_computing_2015}
Rolland J, Bouchet F and Simonnet E 2015 Computing transition rates for the
  1-{D} stochastic {Ginzburg}-{Landau}-{Allen}-{Cahn} equation for
  finite-amplitude noise with a rare event algorithm

\bibitem{rolland_statistical_2015}
Rolland J and Simonnet E 2015 {\em Journal of Computational Physics\/} {\bf
  283} 541--558 ISSN 0021-9991

\bibitem{heymann2008geometric}
Heymann M and Vanden-Eijnden E 2008 {\em Communications on pure and applied
  mathematics\/} {\bf 61} 1052--1117

\bibitem{del_moral_genealogical_2005}
Del~Moral P and Garnier J 2005 {\em The Annals of Applied Probability\/} {\bf
  15} 2496--2534 ISSN 1050-5164

\bibitem{garnier_simulations_2006}
Garnier J and Moral P~D 2006 {\em Optics Communications\/} {\bf 267} 205--214
  ISSN 0030-4018

\bibitem{hairer_improved_2014}
Hairer M and Weare J 2014 {\em Communications on Pure and Applied
  Mathematics\/} {\bf 67} 1995--2021 ISSN 1097-0312

\bibitem{tailleur_probing_2007}
Tailleur J and Kurchan J 2007 {\em Nature Physics\/} {\bf 3} 203--207 ISSN
  1745-2473 wOS:000244558400021

\bibitem{laffargue_large_2013}
Laffargue T, Lam K~D~N~T, Kurchan J and Tailleur J 2013 {\em Journal of Physics
  A: Mathematical and Theoretical\/} {\bf 46} 254002 ISSN 1751-8121

\bibitem{giardina_simulating_2011}
Giardin\`a C, Kurchan J, Lecomte V and Tailleur J 2011 {\em Journal of
  Statistical Physics\/} {\bf 145} 787--811 ISSN 0022-4715, 1572-9613

\bibitem{lecomte_numerical_2007}
Lecomte V and Tailleur J 2007 {\em Journal of Statistical Mechanics: Theory and
  Experiment\/} {\bf 2007} P03004 ISSN 1742-5468

\bibitem{giardina_direct_2006}
Giardin\`a C, Kurchan J and Peliti L 2006 {\em Physical Review Letters\/} {\bf
  96} 120603

\bibitem{kurchan_large_2015}
Kurchan J 2015 {\em Physica A: Statistical Mechanics and its Applications\/}
  {\bf 418} 170--188 ISSN 0378-4371

\bibitem{vanden2009exact}
Vanden-Eijnden E and Venturoli M 2009 {\em The Journal of chemical physics\/}
  {\bf 131} 044120

\bibitem{adams_harmonic_2008}
Adams D~A, Sander L~M and Ziff R~M 2008 {\em Physical Review Letters\/} {\bf
  101} 144102

\bibitem{chandler_interfaces_2005}
Chandler D 2005 {\em Nature\/} {\bf 437} 640--647 ISSN 0028-0836

\bibitem{noe_constructing_2009}
No\'e F, Sch\"utte C, Vanden-Eijnden E, Reich L and Weikl T~R 2009 {\em
  Proceedings of the National Academy of Sciences\/} {\bf 106} 19011--19016
  ISSN 0027-8424, 1091-6490

\bibitem{metzner_illustration_2006}
Metzner P, Sch\"utte C and Vanden-Eijnden E 2006 {\em The Journal of Chemical
  Physics\/} {\bf 125} 084110 ISSN 0021-9606, 1089-7690

\bibitem{bolhuis_transition_2002}
Bolhuis P~G, Chandler D, Dellago C and Geissler P~L 2002 {\em Annual Review of
  Physical Chemistry\/} {\bf 53} 291--318

\bibitem{wolde_enhancement_1997}
Wolde P~R~t and Frenkel D 1997 {\em Science\/} {\bf 277} 1975--1978 ISSN
  0036-8075, 1095-9203

\bibitem{e_energy_2003}
E W, Ren W and Vanden-Eijnden E 2003 {\em Journal of Applied Physics\/} {\bf
  93} 2275--2282 ISSN 0021-8979, 1089-7550

\bibitem{kohn_magnetic_2005}
Kohn R~V, Reznikoff M~G and Vanden-Eijnden E 2005 {\em Journal of Nonlinear
  Science\/} {\bf 15} 223--253 ISSN 0938-8974, 1432-1467

\bibitem{grafke2013instanton}
Grafke T, Grauer R and Sch{\"a}fer T 2013 {\em Journal of Physics A:
  Mathematical and Theoretical\/} {\bf 46} 062002

\bibitem{grafke2014arclength}
Grafke T, Grauer R, Schäfer T and Vanden-Eijnden E 2014 {\em Multiscale
  Modeling \& Simulation\/} {\bf 12} 566--580

\bibitem{lorenz_predictability:_1996}
Lorenz E~N 1996 Predictability: {A} problem partly solved {\em {GARP}
  {Publication} {Series}\/} vol~16 (Geneva, Switzerland: WMO) pp 132--136

\bibitem{lorenz_designing_2005}
Lorenz E~N 2005 {\em Journal of the Atmospheric Sciences\/} {\bf 62} 1574--1587
  ISSN 0022-4928, 1520-0469

\bibitem{vanden-eijnden_rare_2012}
Vanden-Eijnden E and Weare J 2012 {\em Communications on Pure and Applied
  Mathematics\/} {\bf 65} 1770--1803 ISSN 1097-0312

\bibitem{el_makrini_diffusion_2007}
El~Makrini M, Jourdain B and Leli\`evre T 2007 {\em ESAIM: Mathematical
  Modelling and Numerical Analysis\/} {\bf 41} 189--213

\bibitem{del_moral_adaptive_2012}
Del~Moral P, Doucet A and Jasra A 2012 {\em Bernoulli\/} {\bf 18} 252--278 ISSN
  1350-7265

\end{thebibliography}

\end{document}